%% file: arxiv_merged.tex
\documentclass[twocolumn,secnumarabic,amssymb, nobibnotes, aps, prl]{revtex4-2}

\usepackage{amsmath}
\usepackage{bm}
\usepackage{graphicx}
\usepackage{mathbbol}

\usepackage[colorlinks=true, allcolors=blue]{hyperref}

\newcommand{\sps}{\textsl{\textsc{s}}}
\newcommand{\su}{\textsl{\textsc{u}}}
\newcommand{\sv}{\textsl{\textsc{v}}}
\newcommand{\sw}{\textsl{\textsc{w}}}
\newcommand{\sa}{\textsl{\textsc{a}}}
\newcommand{\spb}{\textsl{\textsc{b}}}
\newcommand{\spt}{\textsl{\textsc{t}}}

\newcommand{\figpath}{.}

\begin{document}

\title{Nonreciprocity as a Generic Mechanism for Demixing in Flocking Mixtures}

\author{Charlotte Myin}
\email{charlotte.myin@ds.mpg.de}
\affiliation{Max Planck Institute for Dynamics and Self-Organization (MPI-DS), 37077 Göttingen, Germany}
\author{Benoît Mahault}%
\altaffiliation[Current affiliation:]{~Laboratoire Charles Coulomb (L2C), Univ. Montpellier, CNRS, Montpellier, France}
\email{benoit.mahault@umontpellier.fr}
\affiliation{Max Planck Institute for Dynamics and Self-Organization (MPI-DS), 37077 Göttingen, Germany}
\date{\today}%

\begin{abstract}
We show that even weak nonreciprocal alignment leads to large-scale structure formation in flocking mixtures.
By combining numerical simulations of a binary Vicsek model and the analysis of coarse-grained continuum equations,
we demonstrate that nonreciprocity destabilizes the ordered phase formed by mutually aligning or anti-aligning species in a large part of the phase diagram.  
For aligning populations, this instability results in one species condensing in a single band that travels within a homogeneous liquid of the other species. When interactions are anti-aligning, both species self-assemble into polar clusters with large-scale chaotic dynamics. 
In both cases, the emergence of structures is accompanied by the demixing of the two species, despite the absence of repulsive interactions. Our theoretical analysis allows us to elucidate the origin of the instability, and show that it is generic to nonreciprocal flocks.
\end{abstract}
\maketitle

In recent years, nonreciprocal interactions have been recognized as an essential feature of systems that evolve far from equilibrium, such as active matter~\cite{BowickPRX2022,haluts2024active,golestanian2024non-reciprocal}. Nonreciprocity is particularly relevant to describe biological populations competing for resources through complex social interactions~\cite{Laffertyscience2015,haluts2024active,blumenthal2024phase},
while it arises naturally in simpler contexts, 
including chemotactic mixtures~\cite{soto2014self,meredith2020predator,agudo2019active,duan2023dynamical,dinelli2023non,liu2024self,duanPRR2025} and heterogeneous populations of robots~\cite{chen2024emergent} or colloidal rollers~\cite{colloidalFlocks}. 
One of the most spectacular consequences of nonreciprocity is the emergence of collectively induced dynamical states~\cite{saha2020scalar,you2020nonreciprocity,Fruchart2021Apr}.

Since the seminal works of Vicsek \textit{et al.}~\cite{Vicsek}, Toner and Tu~\cite{TonerPRL95,tonerPRE98}, the study of flocking agents has become a paradigmatic example of active matter~\cite{DADAM,solon2024thirty}.
Nonreciprocity can be introduced into flocking models by limiting the particles’ range of perception, so that they interact, e.g., mainly with neighbors in front of them. This geometric nonreciprocity significantly affects the collective behavior of flocks, leading to the formation of structures such as condensates~\cite{chenPRE2017}, lanes, or vortex states~\cite{minimalcognflockingmodel, milling, visualpercept}.
Another natural way in which nonreciprocity arises is when heterogeneous populations interact asymmetrically~\cite{YllanesNJP2017,Fruchart2021Apr,KreienkampNJP2022,KreienkampPRL2024,chen2024emergent,TucciNJP2024,kreienkampComPhys2025,TangPRR2025,mihatsch2025nonreciprocal,martin2025transition,Scita2025}. In this case, nonreciprocity can induce dynamical chiral states through the crossing of exceptional points~\cite{Fruchart2021Apr,zellePRX2024,kreienkampComPhys2025}. Importantly, such collective behaviors require {\it strongly} nonreciprocal interactions, resulting in antisymmetric couplings between the two populations.

In this Letter, we focus on a comparatively less explored regime, and investigate the fate of flocks subjected to weak population nonreciprocity. 
Considering a Vicsek model (VM) of two populations that mutually align or anti-align their velocities, we demonstrate that nonreciprocal interactions generally induce large-scale structure formation associated with the demixing of the two species. 
Unlike in previous studies, the mechanism underlying demixing solely relies on nonreciprocal alignment and self-propulsion, and does not require additional features such as population heterogeneity~\cite{colloidalFlocks} or pairwise repulsion~\cite{KreienkampPRL2024,KreienkampPRE2024}.
We further show, using a coarse-grained continuous model, that this phenomenology originates from a generic long-wave instability, which we trace back to the coupling between density and orientational order inherent to the physics of flocks~\cite{DADAM}.

\indent\textit{Microscopic model.}---The binary VM consists of point-like self-propelled particles in two dimensions, 
aligning their direction of motion with neighbors within unit distance. 
The position $\bm r_k^{\sps}$ and orientation $\theta_k^{\sps}$ of particle $k$ 
%$(k = 1, \dots, N^{\sps})$
from species $\sps = \sa,\spb$ evolve in discrete time according to 
\begin{subequations}
\begin{align}
	\label{micromodel_r}
  	\bm r_k^{\sps}(t + 1) & = \bm r_k^{\sps}(t) + v_0^{\sps} \hat{\bm u}[\theta_k^{\sps}(t+1)],\\
	\label{micromodel_t}
    	\theta_k^{\sps}(t + 1) & = {\rm arg}\bigg[ \sum_{\su, j} \chi^{\sps \su} \, e^{i \theta_j^{\su}(t)} + \eta^{\sps} n_k(t) e^{i \xi_k(t)} \bigg] . 
\end{align}
\label{micromodel}
\end{subequations}
Here, $v_0^{\sps}$ and $\eta^{\sps}$ denote the self-propulsion speed and noise strength associated with species $\sps$, respectively.
$\hat{\bm u}(\theta) = \begin{pmatrix} \cos\theta & \sin\theta\end{pmatrix}^T$ and $\xi_k$ is a delta-correlated white noise uniformly drawn from $[-\pi,\pi]$, 
while $n_k(t)$ is the total number of interacting neighbours of particle $k$ (including itself) at time $t$. 

The sum in Eq.~\eqref{micromodel_t} runs over $\su = \sa, \spb$ and particles $j$ satisfying $|\bm r_k^{\sps} - \bm r_j^{\su}| \le 1$, while the coupling $-1 \le \chi^{\sps\su} \le 1$ defines how particles from species $\sps$ align ($\chi^{\sps\su} > 0$) or anti-align ($\chi^{\sps\su} < 0$) their velocities with particles from species $\su$.
The average coupling $\bar\chi \equiv \frac{1}{2}(\chi^{\sa\spb} + \chi^{\spb\sa})$ then describes the tendency of the two species to mutually align ($\bar\chi > 0$) or anti-align ($\bar\chi < 0$) their velocities.
In turn, enforcing reciprocal interspecies alignment imposes that $\Delta \chi \equiv \frac{1}{2}(\chi^{\sa\spb} - \chi^{\spb\sa}) = 0$,
while interactions are nonreciprocal otherwise.

When $\chi^{\sps\su} = 1$ for all $\sps,\su \in \{\sa,\spb\}$,
the model reduces to the classical VM, 
while other limiting cases such as 
fully anti-aligning mixtures~\cite{kurstenPRE2025},
reciprocally anti-aligning populations ($\chi^{\sa\sa,\spb\spb} > 0$, $\bar\chi < 0$, $\Delta\chi = 0$)~\cite{MenzelPRE2012,MahaultPRL2021,unfriendlyspecies}
and self-anti-aligning species ($\chi^{\sa\sa,\spb\spb} < 0$) with aligning interspecies interactions ($\bar\chi > 0$)~\cite{Lardet2025Mar,Oki2025Feb} have been addressed in the literature. 
Here, we let particles of the same species align ($\chi^{\sa\sa,\spb\spb}  =1$) 
while varying the interspecies couplings $\chi^{\sa\spb}$ and $\chi^{\spb\sa}$.
In the regime of strong nonreciprocity, where $|\Delta\chi| > |\bar\chi|$, we observe oscillatory behavior consistently with results reported in the literature~\cite{Fruchart2021Apr,kreienkampComPhys2025,martin2025transition}.
Hereafter, however, we focus on the weakly nonreciprocal regime, where $|\Delta\chi| \le |\bar\chi|$.

\begin{figure}
    \centering
    \includegraphics[width=\columnwidth]{\figpath/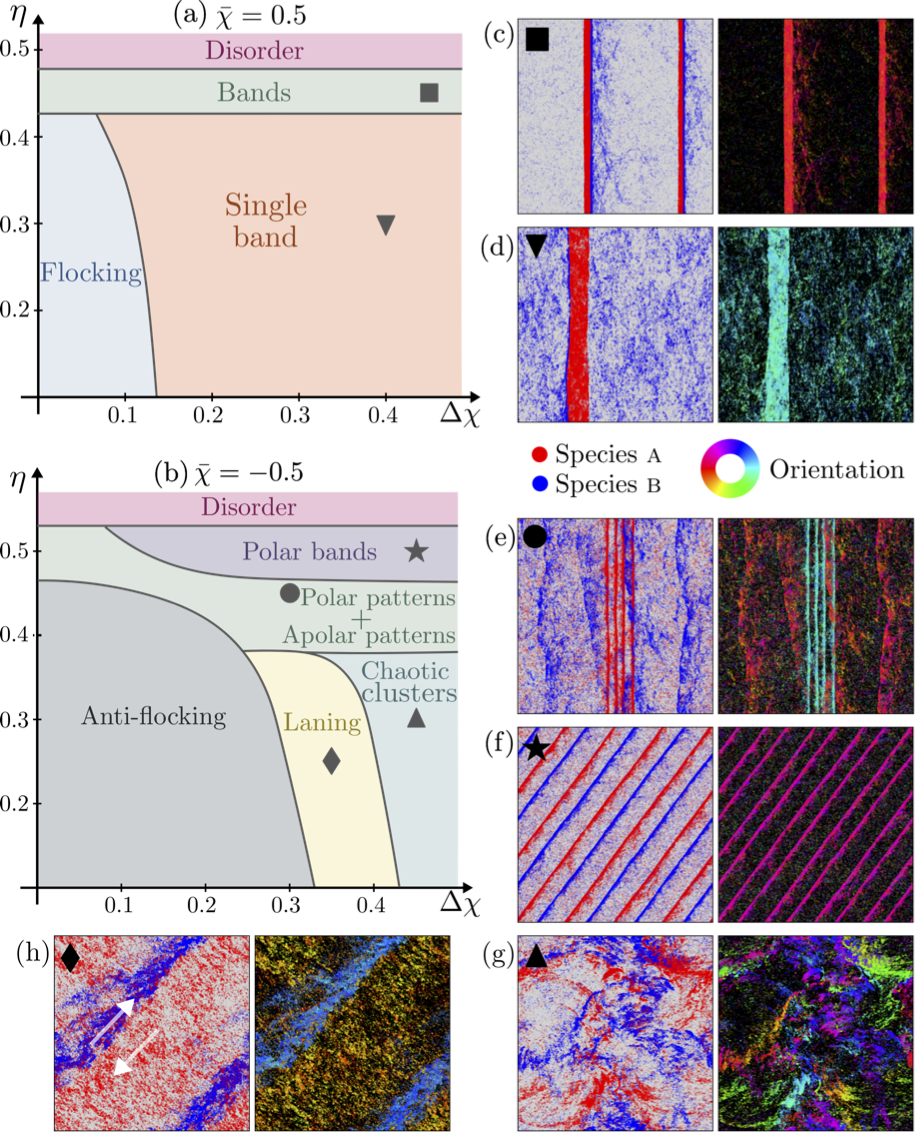}
    \caption{Phase behaviour of weakly nonreciprocal flocks.
    (a,b) Stylized finite-size phase diagrams in the $(\Delta\chi,\eta)$ plane for aligning (a) and anti-aligning (b) intraspecies interactions. 
    (c)-(h) Representative snapshots of the observed phases. 
    Particles are colour-coded according to their species (orientation) in the left (right) panel (caption below (d)),
    while the corresponding points in (a,b) are indicated with solid symbols.
    Additional details on the construction of (a,b) and simulation parameters are provided in~\cite{supplement}.
    }
    \label{fig:fig1}
\end{figure}

As we focus on the effects induced by nonreciprocity alone, 
we discard the possible influence of population heterogeneity~\cite{DuttaPRE2025} by setting $\eta^{\sa,\spb} = \eta$ and $v_0^{\sa,\spb} = v_0$~\footnote{We show in~\cite{supplement} that our conclusions remain unaffected when introducing population heterogeneity.}.
We also consider equal densities $\rho^{\sa,\spb} = \rho_0 = N/L^2$, 
where $N$ denotes the size of each population and $L$ is the linear size of the square domain.
Unless specified otherwise, we use $\rho_0 = \tfrac{1}{2}$, $v_0 = 1$, $\bar\chi = \pm \tfrac{1}{2}$,
and vary the angular noise $\eta$ and the nonreciprocity parameter $\Delta \chi$.
Noting that the problem is symmetric under the exchange $\sa \leftrightarrow \spb$, 
we further restrict the analysis to $\Delta\chi \ge 0$.
To construct the phase diagrams of Fig.~\ref{fig:fig1}(a,b), we numerically
integrated Eqs.~\eqref{micromodel} in periodic square domains of size $L = 512$.
Regardless of the sign of $\bar\chi$, the system is disordered at high noise and transitions to a symmetry broken phase with nonzero global polar orders $\bm \Pi^{\sps} = \langle \hat{\bm u}(\theta^{\sps}_k) \rangle_{k}$ when $\eta$ decreases. 
For all values of $\Delta\chi$, this transition is mediated by a phase-coexistence region, 
reminiscent of the microphase separation scenario at play in the single-species VM~\cite{gregoirePRL2004,Microphase}.

%%%%%%%%%%%%%%%%%%%%%%%%%%%%%%%%%%%
%%%%%%%%%%%%%%%%%%%%%%%%%%%%%%%%%%%

\textit{Mutually aligning populations.}---
For $\bar\chi >0$ at the onset of order, both species self-organize into dense bands propagating through a dilute, gaseous background.
As shown in Fig.~\ref{fig:fig1}(c) and in the Supplementary Material~\cite{supplement}, 
bands are mostly composed of a single species and are systematically arranged in pairs, with particles having the lower interspecies coupling (here $\spb$, in blue) at the rear.
Such spatial synchronization by nonreciprocal alignment---also reported for discrete symmetry flocks~\cite{martin2025transition}---can be qualitatively understood from a simple argument detailed in \hyperlink{appA}{Appendix A}.

\begin{figure}
    \centering
    \includegraphics[width=\columnwidth]{\figpath/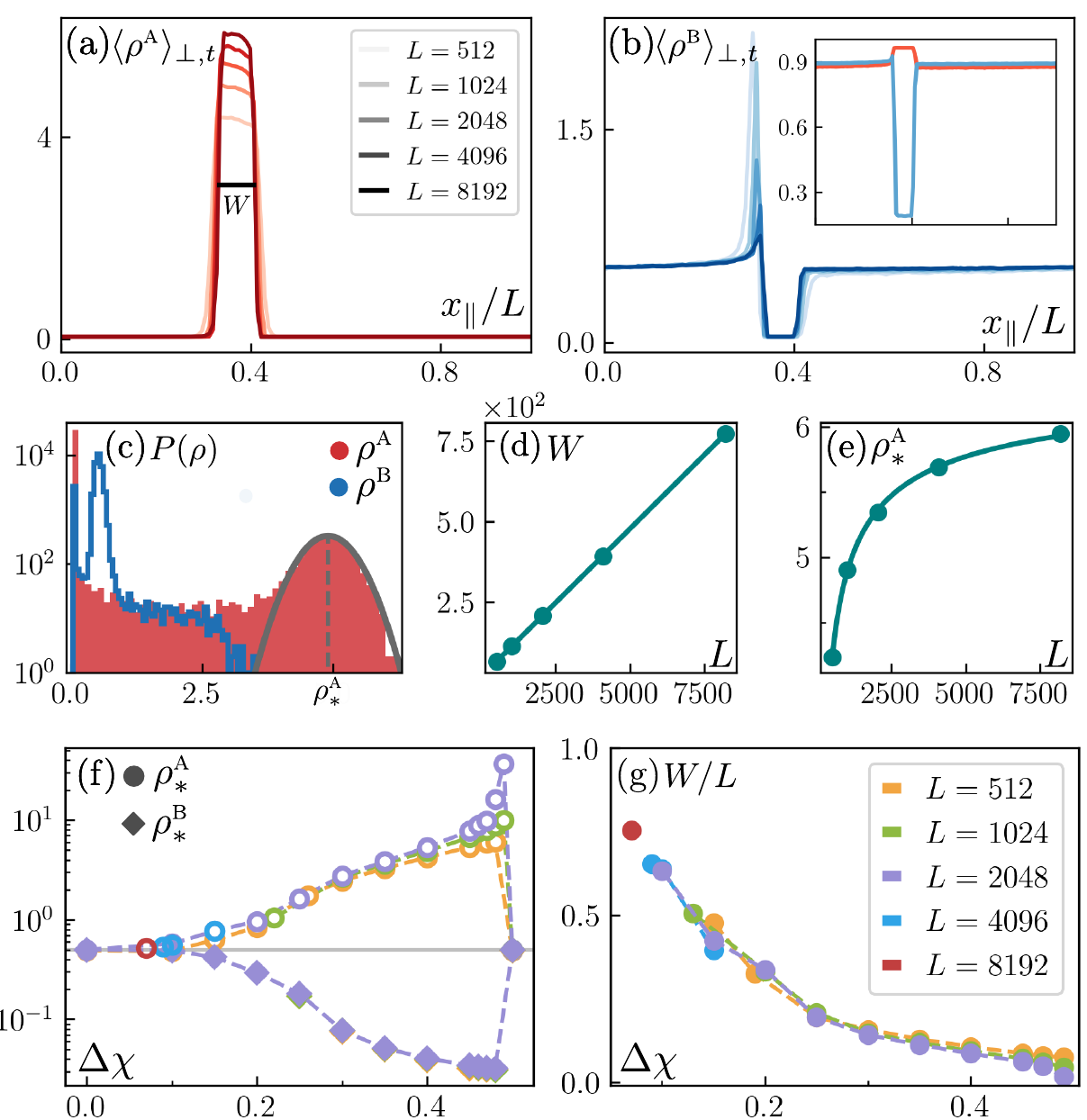}
    \caption{Characterisation of the single band phase. 
    (a,b) $\rho^{\sa}$(a) and $\rho^{\spb}$(b) profiles averaged along the band and over time for different system sizes. 
    Inset of (b): the component of the local polarity profiles along the global order.
    (c) Density histograms at $L = 1024$. $\rho^{\sa}_{*}$ is shown with a vertical dashed line, while $\rho^{\spb}_{*}$ is defined as the position of left peak of the blue histogram.
    (d,e) Band width (d) and density (e) as functions of system size.
    Symbols show the numerical data and solid lines indicate fits (see text).
    (f,g) The densities in the band (f) and its normalized width (g) as functions of $\Delta\chi$ for various system sizes. 
    The horizontal grey line in (f) indicates $\rho_0$.
    In all panels, $\eta =0.3$ and for (a-e) $\Delta\chi = 0.4$.}
    \label{fig:fig2}
\end{figure}

Deeper in the ordered phase and for sufficiently low $\Delta \chi$,
the species form a homogeneous flock, characterized by aligned mean polarities $\langle\bm \Pi^{\sa,\spb}\rangle_t$. 
However, this phase becomes unstable for sufficiently high $\Delta\chi$, where nonreciprocal alignment gives rise to traveling bands primarily made of $\sa$ particles, which here align more strongly.
After a coarsening process, the steady-state eventually consists of a single band [Fig.~\ref{fig:fig1}(d)].
To characterize this structure, we use density profiles averaged along the band and over time, as depicted in Figs.~\ref{fig:fig2}(a,b).
Contrary to the microphase coexistence region at higher noise, the density profiles reveal that the $\sa$-species band coincides with a local depletion of $\spb$ particles, and moves within a homogenous polar liquid of the $\spb$ species.

To determine the characteristic densities $\rho^{\sa}_*$ and $\rho^{\spb}_*$ within the band, we use the position of the right and left peak of the $\sa$ and $\spb$ density histograms, respectively, as illustrated in Fig.~\ref{fig:fig2}(c).
The band width $W$ is then defined from the half maximum of time-averaged $\sa$ density profiles. Performing simulations in systems of increasing size, we show in Fig.~\ref{fig:fig2}(d) that $W$ grows linearly with $L$.
Moreover, $\rho^{\sa}_*$ exhibits an unusually slow convergence to its asymptotic value,
which we best fit as a power law: $\rho^{\sa}_* \simeq \rho^{\sa}_\infty - A L^{-\nu}$ with $\nu \simeq \tfrac{1}{2}$ [Fig.~\ref{fig:fig2}(e)].
These measures indicate that, 
up to strong finite size effects, 
the single-band phase displays characteristics of a phase-separated state in which both species spontaneously demix,
despite the absence of repulsive interactions. 
Figures~\ref{fig:fig2}(f,g) illustrate how the properties of this phase vary with the amount of nonreciprocity. For small $\Delta\chi$, the width of the band $W$ is comparable to the system size $L$, while it induces weak density modulations. As $L$ grows, the homogeneous flocking phase becomes more susceptible to the presence of nonreciprocity, such that the demixed phase is found for increasingly smaller values of $\Delta\chi$, 
down to $\Delta\chi = 0.07$ for $L = 8192$ ($N = 67\!\cdot\!10^6$ particles).
Despite our computational efforts, we were unable to determine whether the transition from a homogeneous flock to the demixed phase arises at a finite value of $\Delta\chi$, 
or if the latter can survive an infinitesimal amount of nonreciprocity.
Increasing $\Delta\chi$ rapidly enhances demixing [note the vertical logarithmic scale in Fig.~\ref{fig:fig2}(f)], such that for $\Delta\chi \lesssim \bar\chi$,
the densities $\rho^{\sa,\spb}_*$ are orders of magnitude apart and $W$ takes values close to zero. Remarkably, the resulting thin and dense band exists even for values of $\chi^{\spb\sa} = \bar\chi - \Delta\chi$ as low as $10^{-2}$,
while it naturally disappears when $\Delta\chi = \bar\chi$.

%%%%%%%%%%%%%%%%%%%%%%%%%%%%%%%%%%%
%%%%%%%%%%%%%%%%%%%%%%%%%%%%%%%%%%%

\textit{Mutually anti-aligning populations.}---
For $\bar\chi < 0$ and low nonreciprocity,
the coexistence phase at the onset of order may consist of 
trains of bands moving in opposite directions [Fig.~\ref{fig:fig1}(e)]. 
In these apolar arrangements, bands appear in equal number and are regularly spaced for $\Delta\chi = 0$~\cite{supplement}, while their structure becomes irregular at higher nonreciprocity.
Consistently with results reported for $\Delta\chi = 0$~\cite{unfriendlyspecies}, these configurations coexist in a large part of the coexistence phase with polar arrangements of bands of alternating species that travel in the same direction [Fig.~\ref{fig:fig1}(f)]. 
We do not observe dynamical transitions between polar and apolar patterns, which are selected by the initial condition.
Apolar configurations occur more frequently at low nonreciprocity and noise,
while increasing $\Delta\chi$, 
polar trains arise more often~\cite{supplement}.
In fact, the latter become the only observed configuration for sufficiently large $\eta$ and $\Delta\chi$ [purple region in Fig.~\ref{fig:fig1}(b)].

Similarly to aligning populations,
the homogeneous liquid with anti-aligned mean polar orders $\langle\bm \Pi^{\sa,\spb}\rangle_t$ found at $\Delta\chi = 0$ is only stable up to a finite amount of nonreciprocity. 
Beyond a $\Delta\chi$ threshold, the two species self-organize in a laning pattern consisting of alternating domains with opposite polarities 
[Fig.~\ref{fig:fig1}(h) and Supplementary movie SMov~1]. 
This pattern is highly dynamic, as the lanes persistently rotate and drift over long timescales (SMov~2),
while they also intermittently break and reform.
Increasing $\Delta\chi$,
breaking events become more frequent,
eventually preventing the pattern from repairing itself.
At sufficiently high $\Delta\chi$, the populations demix by arranging into polar clusters mainly made of a single species, 
which move in a chaotic manner and lack macroscopic spatial order [Fig. \ref{fig:fig1}(g) and SMov~3].
The lane breaking events, which are associated with a sharp decay of polar order for both species [see Fig.~\ref{fig:figappA}(a) in \hyperlink{appB}{Appendix B}], happen more often in larger systems (SMov~4).
As a result, the system becomes progressively disordered as revealed by the distributions of $|\bm \Pi^{\sps}|$ shown in Fig.~\ref{fig:figappA}(b).
We therefore expect the laning pattern to eventually disappear in the thermodynamic limit,
resulting in a disordered polar cluster phase.

\textit{Hydrodynamic theory}---To gain a theoretical understanding of the above findings,   
we coarse-grained the microscopic model~\eqref{micromodel} into a continuous theory.
Because of its success at qualitatively capturing the physics of the single-species VM, we followed the Boltzmann-Ginzburg-Landau approach~\cite{bertinPRE2006,bertinjpa2009,Peshkov2014Jun}.
As detailed in \hyperlink{appC}{Appendix C}, we first write the Boltzmann equation governing the evolution of the single-body distribution function $f^{\sps}(\bm r,\theta,t)$ associated with species $\sps$.
By expanding $f^{\sps}$ in angular Fourier modes, 
we then obtain
infinite coupled hierarchies of nonlinear equations. 
The zeroth and first modes correspond to the local particle densities ($\rho^{\sa,\spb}$) and polarities ($\bm p^{\sa,\spb}$), respectively, which are the physical fields of interest. 
Using a standard scaling ansatz describing the relative amplitude of the modes and their spatio-temporal variations at the onset of order, we truncate and close the hierarchies at the first nontrivial order.
The resulting hydrodynamic equations for $\rho^{\sps}$ and $\bm p^{\sps}$ read
\begin{subequations}\label{hydro}
\begin{align}
\label{hydro_rho}
        \partial_t \rho^{\sps} & + v_0\nabla \cdot \bm{p}^{\sps} = 0, \\
    \partial_t p^{\sps}_i & +
    \sum_{\su,\sv
}\left[\psi_{\su\sv}^{\sps} (\bm p^{\sv}\cdot\nabla) p^{\su}_i
    + \lambda_{\su\sv}^{\sps} (\nabla\cdot\bm p^{\sv})p^{\su}_i
    + \nu_{\su\sv}^{\sps} \bm p^{\su} \cdot \partial_i \bm p^{\sv} \right]\nonumber \\
    & = \sum_{\su}
    \left[\alpha^{\sps}_{\su}(\rho^{\sa},\rho^{\spb}) 
    - \sum_{\sv,\sw
    }\xi^{\sps}_{\su\sv\sw} (\bm p^{\sv}\cdot \bm p^{\sw}) \right]p^{\su}_i
    -\frac{v_0}{2}\partial_i\rho^{\sps} \nonumber \\
    \label{hydro_p}
    & + \sum_{\su
    }D^{\sps}_{\su}\nabla^2 p^{\su}_i. 
\end{align} 
\end{subequations}
Except for $\alpha^{\sps}_{\su}$ which varies linearly with the densities,
all coefficients in Eqs.~\eqref{hydro} are constant. In~\cite{supplement}, we provide their formal expressions in terms of integrals involving the microscopic model parameters, as well as a Mathematica notebook for computing them numerically.

\begin{figure}
    \centering
    \includegraphics[width=\columnwidth]{\figpath/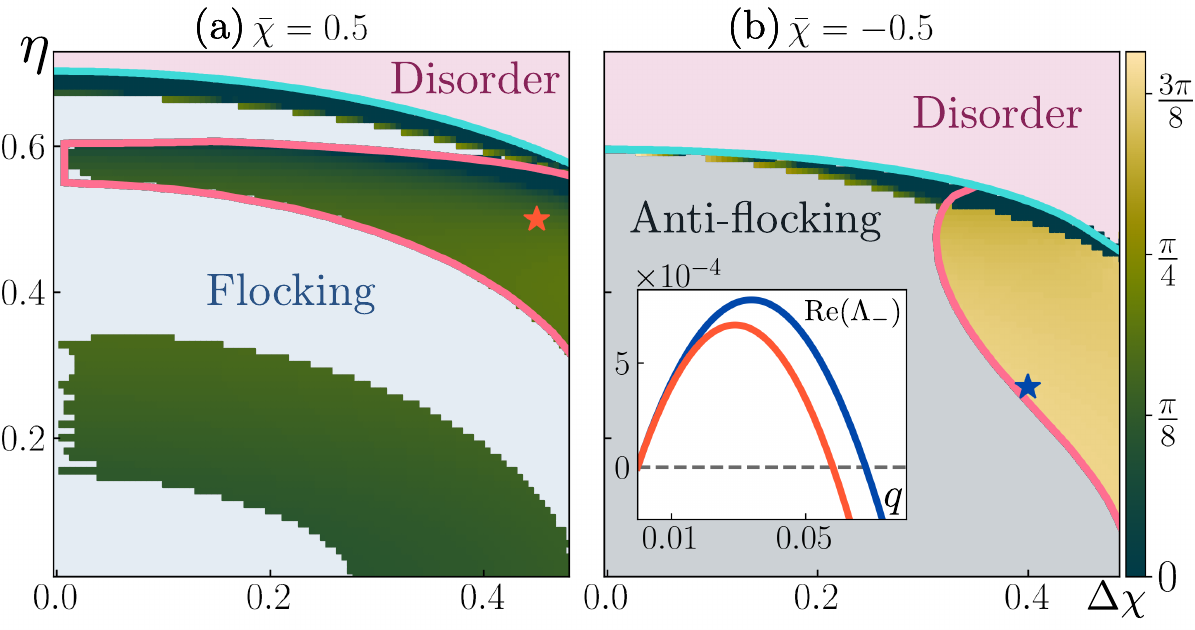}
    \caption{Phase diagrams in the $(\Delta\chi, \eta)$ plane obtained from the linear stability of the stationary homogeneous solutions of Eqs.~\eqref{hydro} in the flocking (a) and anti-flocking (b) regimes.
    Stable disordered, flocking and anti-flocking regions appear in green, blue and gray, respectively.
    The other regions indicate an instability, and are colored according to the orientation of the most unstable wave vector with respect to the direction of order.
    The solid cyan lines mark the onset of order, while the pink lines enclose regions where the matrix $\mathbb{M}_{\rm h}(\bm q)$ in Eq.~\eqref{mat_ensl} predicts an instability. 
    Inset of (b): Growth rate of the flocking (blue) and anti-flocking (red) instabilities as functions of the wave number $q = |\bm q|$ along the most unstable direction.
    }
    \label{fig:fig3}
\end{figure}

Expressing the polarities as $\bm p^{\sps} = s^{\sps} \hat{\bm u}(\theta^{\sps})$, 
we rewrite Eq. \eqref{hydro_p} in terms of their magnitudes $s^{\sps}$ and orientations $\theta^{\sps}$. 
The homogeneous, stationary solutions of Eqs.~\eqref{hydro} correspond to uniform $\rho^{\sps} = \rho^{\sps}_0$ and $s^{\sps} = s_0^{\sps}$, which solve for $\su \ne \sps$
\begin{align}
    \label{homsol}
    F^{\sps}_{\sps}s_0^\sps + F^{\sps}_{\su}s_0^\su \cos(\Delta\theta_{\sps\su}) = 0 , 
    \quad
    F^{\sps}_{\su}s_0^{\su}\sin(\Delta\theta_{\sps\su}) = 0,
\end{align}
where $F^{\sps}_{\su} \equiv \alpha^{\sps}_{\su}(\rho_0^{\sa},\rho_0^{\spb}) - 
    \sum_{\sv,\sw}\xi^{\sps}_{\su\sv\sw} s_0^\sv s_0^\sw \cos(\Delta\theta_{\sv\sw})$ 
    and $\Delta\theta_{\sps\su} \equiv \theta^\sps- \theta^\su$. 
    When $|\Delta\chi| < |\bar\chi|$, Eqs.~\eqref{homsol} admit three types of solutions. 
    For sufficiently high noise, $s_0^{\sa,\spb} = 0$ indicating a disordered state.
    Decreasing $\eta$ below a threshold value marked by the cyan lines in Figs.~\ref{fig:fig3}, the solutions bifurcate to an ordered state with $s^{\sa,\spb}_0 >0$, 
    corresponding to a flocking ($\Delta \theta_{\sa\spb} = 0$, $\bar\chi > 0$) or anti-flocking ($\Delta \theta_{\sa\spb} = \pi$, $\bar\chi < 0$) configuration.

    To analyse the stability of these solutions, we linearise Eqs.~\eqref{hydro} for small perturbations, which we express in Fourier space:
    $(\delta\rho^{\sps}, \delta\theta^{\sps}, \delta s^{\sps})(\bm r,t) \to (\delta\hat{\rho}^{\sps}, \delta\hat{\theta}^{\sps}, \delta\hat{s}^{\sps})(\bm q,t)$. 
    We obtain a six-dimensional linear system $\partial_t\bm \hat{\bm X} = \mathbb{M}(\bm q)\bm \hat{\bm X}$, 
    where $\hat{\bm X} \equiv \begin{pmatrix} \delta \hat\rho^\sa & \delta\hat\rho^\spb & \delta\hat\theta^\sa & \delta\hat\theta^\spb &
    \delta \hat{s}^\sa & 
    \delta \hat{s}^\spb \end{pmatrix}^T$ and the matrix $\mathbb{M}(\bm q)$ is given in~\cite{supplement} for completeness. 
    We construct the linear stability diagrams for both $\bar \chi >0$ [Fig. \ref{fig:fig3}(a)] and $\bar\chi <0$ [Fig. \ref{fig:fig3}(b)] by numerically evaluating the eigenvalues of $\mathbb{M}(\bm q)$ for various values of $\eta$ and $\Delta\chi$, and wave vectors $\bm q$.    
    In qualitative agreement with the VM phenomenology~\cite{DADAM} and our microscopic model simulations, both the flocking and anti-flocking solutions are unstable at the onset of order. 
    Similarly to single species flocks, the resulting long-wavelength instability is characterized by a growth rate ${\rm Re}(\Lambda) {\simeq} |\bm q|^2$ for $|\bm q|\to 0$~\cite{bertinPRE2006,ChateLectureNotes}.
    We uncover an additional long-wavelength instability, located deeper in the ordered phase, and which typically requires sufficiently high nonreciprocity.
    For both the flocking and anti-flocking cases this instability exhibits a distinct scaling: ${\rm Re}(\Lambda) {\simeq} |\bm q|$ as $|\bm q|\to 0$ [inset of Fig.~\ref{fig:fig3}(b)], indicating that it originates from a new mechanism~\footnote{For $\bar\chi > 0$, Fig.~\ref{fig:fig3}(a) displays a third unstable region at low noises. As Eqs.~\eqref{hydro} are derived perturbatively at the onset of order, this instability arises far from the formal range of validity of the continuum model and, like for single species flocks, we believe that it is an artifact resulting from approximations made during the coarse-graining~\cite{supplement,Peshkov2014Jun}.}.

%%%%%%%%%%%%%%%%%%%%%%%%%%%%%%%%%%
%%%%%%%%%%%%%%%%%%%%%%%%%%%%%%%%%%

To elucidate its origin, we consider the limit $|\bm q| \to 0$ where non-hydrodynamic modes can be enslaved.
The hydrodynamic modes ---for which $\partial_t \hat{(\cdot)} = \mathcal{O}(|\bm q|)$--- correspond to the left eigenvectors of $\mathbb{M}(\bm 0)$ with zero eigenvalue. 
    In the (anti-)flocking region, we find three of them: the two density fields $\delta\hat{\rho}^{\sa,\spb}$, 
and 
$\bar{\theta} = s^\sa_0 F^{\spb}_{\sa}\delta \hat{\theta}^\sa + {s^\spb_0}{F^{\sa}_{\spb}}\delta\hat{\theta}^\spb$. 
Enslaving the fast modes and keeping only leading order terms in $|\bm q|$ (details in \hyperlink{appD}{Appendix D}), 
the dynamics reduces to a three-dimensional linear system $\partial_t \hat{\bm X}_{\rm h} = \mathbb{M}_{\rm h}(\bm q) \hat{\bm X}_{\rm h}$, 
where $\hat{\bm X}_{\rm h} = \begin{pmatrix}\delta\hat{\rho}^\sa & \delta\hat{\rho}^\spb & \bar\theta\end{pmatrix}^T$, and
\begin{align}\label{mat_ensl}
    \mathbb{M}_{\rm h}(\bm q) = 
        i\begin{pmatrix}
        V_{\sa\sa} q_\| & V_{\sa\spb} q_\| & V_{\sa\theta} q_\perp \\
        V_{\spb\sa} q_\| & V_{\spb\spb} q_\| & V_{\spb\theta} q_\perp \\
        V_{\theta\sa} q_\perp & V_{\theta\spb} q_\perp & V_{\theta\theta} q_\|
    \end{pmatrix} + \mathcal{O}(|\bm q|^2).
\end{align} 
$q_\|$ and $q_\perp$ in Eq.~\eqref{mat_ensl} denote the components of $\bm q$ along and transverse to order, respectively,
while the $V$ terms depend on the coefficients of Eq.~\eqref{hydro} in a complex way~\cite{supplement}.
To further simplify the analysis, we note from Fig.~\ref{fig:fig3}(a) that upon decreasing $\eta$ the homogeneous flocking solution is firstly unstable to longitudinal perturbations, such that we set $q_\perp = 0$.
The matrix $\mathbb{M}_{\rm h}$ then has a purely imaginary eigenvalue $\Lambda_\theta = i V_{\theta\theta}q_\|$,
while the remaining two are given by
\begin{align} \label{Lambdapm}
    \Lambda_{\pm} = \frac{i q_\|}{2}\left( V_{\sa\sa} + V_{\spb\spb} \pm \sqrt{\cal R} \right),
\end{align}
where $\mathcal{R} \equiv 4 V_{\sa\spb} V_{\spb\sa} + (V_{\sa\sa} - V_{\spb\spb})^2$.
When $\mathcal{R} > 0$, the eigenvalues in Eq.~\eqref{Lambdapm} are purely imaginary, and the stability of the solution to long-wavelength perturbations is determined at ${\cal O}(|\bm q|^2)$.
On the other hand, for $\mathcal{R} < 0$ ---i.e., when $V_{\sa\spb} V_{\spb\sa} < -\tfrac{1}{4} (V_{\sa\sa} - V_{\spb\spb})^2$---
the eigenvalue $\Lambda_-$ acquires a positive real part, 
indicating the emergence of an instability. 
For reciprocally aligning populations $V_{\sa\spb}$ and $V_{\spb\sa}$ share the same sign and ${\cal R} > 0$,
while the condition $V_{\sa\spb} V_{\spb\sa} < 0$ requires a sufficient amount of nonreciprocity.
Anti-flocking mixtures are more difficult to analyze, as in this case the instability is not purely longitudinal or transverse [Fig.~\ref{fig:fig3}(b)].
Nevertheless, the solid pink lines in Figs.~\ref{fig:fig3} enclosing regions for which $\mathbb{M}_{\rm h}$ has an eigenvalue with positive real part show excellent agreement with the direct numerical estimate based on the six-dimensional system.
When the instability is longitudinal, these lines correspond to ${\cal R} = 0$. 

It is clear from Eqs.~(\ref{mat_ensl},\ref{Lambdapm}) that the new instability relies on the coupling between the two density fields. 
Therefore, it cannot arise in the single-species Vicsek model, or when nonreciprocally aligning spins are fixed on a lattice~\cite{guislainPRE2024,avni2025nonreciprocal,blomPRE2025}.
This also explains why it was missed in Ref.~\cite{Fruchart2021Apr}, where fluctuations of the densities are discarded in the analysis of the continuum model.
Although $\cal R$ cannot be compactly expressed in terms of the microscopic model parameters,
the density couplings in Eq.~\eqref{mat_ensl} can be written as 
$V_{\sps\su} = v_0^{\sps} \partial_{\rho_\su}{\cal L}^{\sps}$, where ${\cal L}^{\sps}$ is a linear combination of the $\alpha^{\su}_{\sv}(\rho^{\sa},\rho^{\spb})$ coefficients in Eq.~\eqref{hydro_p} (details in \hyperlink{appD}{Appendix D}). 
This formulation notably highlights that the instability is driven by a combination of two ingredients: the particles' self-propulsion encoded by $v_0^{\sps}$, and the local (anti)-alignment of velocities, 
which is reflected in the density dependencies of the $\alpha^{\su}_{\sv}$ coefficients.

We have shown that even a weak amount of nonreciprocity significantly impacts the collective dynamics of flocking active matter. 
Combining agent-based simulations and the analysis of a coarse-grained theory, our results demonstrate that nonreciprocal alignment alone constitutes a generic route to demixing, without resorting to other mechanisms such as population heterogeneity~\cite{colloidalFlocks} or pairwise repulsion~\cite{KreienkampPRL2024,KreienkampPRE2024}.
Our theoretical analysis further emphasises that this phenomenology is induced by the coupling between density and orientational order, which shapes most of the characteristics of flocks.
Our findings echo recent works highlighting the metastability of flocks~\cite{obstacle,BenvegnenPRL2023} and their fragility to the presence of spatial anisotropy~\cite{SolonPRL2022}, inclusions and boundaries~\cite{obstacle,FavaPRL2024}, quench~\cite{morinNatPhys2017,chardacPNAS2021,quenceddisorder} and chirality~\cite{VentejouPRL2021} disorder.  
Since they rely solely on weakly nonreciprocal alignment, we believe that our results will have implications for the modeling of a variety of systems,
including mixtures of Quincke rollers~\cite{colloidalFlocks} or Janus colloids~\cite{TucciNJP2024}, and migrating heterogeneous cellular tissues~\cite{Scita2025}.

\acknowledgements
We thank Hugues Chaté and Yu Duan for their critical reading of our manuscript, as well as Suropriya Saha for insightful discussions. C.M. acknowledges funding from the International Max Planck Research School (IMPRS) for the Physics of Biological and Complex Systems.

\bibliographystyle{apsrev4-2}
\bibliography{sample.bib}

\onecolumngrid

\vspace{12pt}
\noindent\hrulefill \hspace{24pt} {\bf End Matter} \hspace{24pt} \hrulefill
\vspace{12pt}

\twocolumngrid

\renewcommand \thefigure{A\arabic{figure}}
\setcounter{figure}{0}
\setcounter{equation}{0}
\renewcommand{\theequation}{A\arabic{equation}}
\hypertarget{appA}{\textit{Appendix A: Synchronization of Vicsek bands via nonreciprocal alignment.---}} 
Here, we provide a qualitative argument explaining the synchronization of polar bands
such as those shown in Fig.~\ref{fig:fig1}(c).
We consider a pair of $\sa$ and $\spb$ bands traveling with the same velocity $v_{\rm b}^{\sa} = v_{\rm b}^{\spb}$.
Like in Fig.~\ref{fig:fig1}(c), the band mostly composed of the species with the largest interspecies coupling (here $\sa$, in red), is at the front.
Increasing the separation between the two bands, 
particles interact with fewer neighbors of the other species.
As a result, the local polarization in the $\sa$ band decays faster than that of the $\spb$ band (since $\chi^{\sa\spb} > \chi^{\spb\sa}$).
Since the velocity of each band is proportional to the local polar order inside (from Eq.~\eqref{micromodel_r}), 
this leads to $v_{\rm b}^{\sa} < v_{\rm b}^{\spb}$, causing the $\spb$ band to catch up. 
Similarly, reducing the inter-band separation yields $v_{\rm b}^{\sa} > v_{\rm b}^{\spb}$, such that the initial configuration is stabilized by nonreciprocity.
The reversed configuration with the $\spb$ band in front is, in turn, always unstable.

\renewcommand \thefigure{B\arabic{figure}}
\setcounter{figure}{0}
\setcounter{equation}{0}
\renewcommand{\theequation}{B\arabic{equation}}
\hypertarget{appB}{\textit{Appendix B: Stability of the laning pattern.---}} 
As discussed in the main text, when increasing the system size the laning pattern reported in Figs.~\ref{fig:fig1}(b,h) for mutually anti-aligning species becomes more vulnerable to breaking events (see also SMov~4).
Each of these events appears as a sharp decay of the global polar order $|\bm \Pi^{\sps}|$ for both species, as shown in Fig.~\ref{fig:figappA}(a) for $\sps = \sa$. 
Hence, the corresponding polarity distributions develop a fat tail at low order [Fig.~\ref{fig:figappA}(b)] which is more prominent in larger systems, indicating that the latter become progressively disordered.

%%%%%%%%%%%%%%%%%%%%%%%%%%%%%%%%%%
\begin{figure}
    \centering
    \includegraphics[width=\columnwidth]{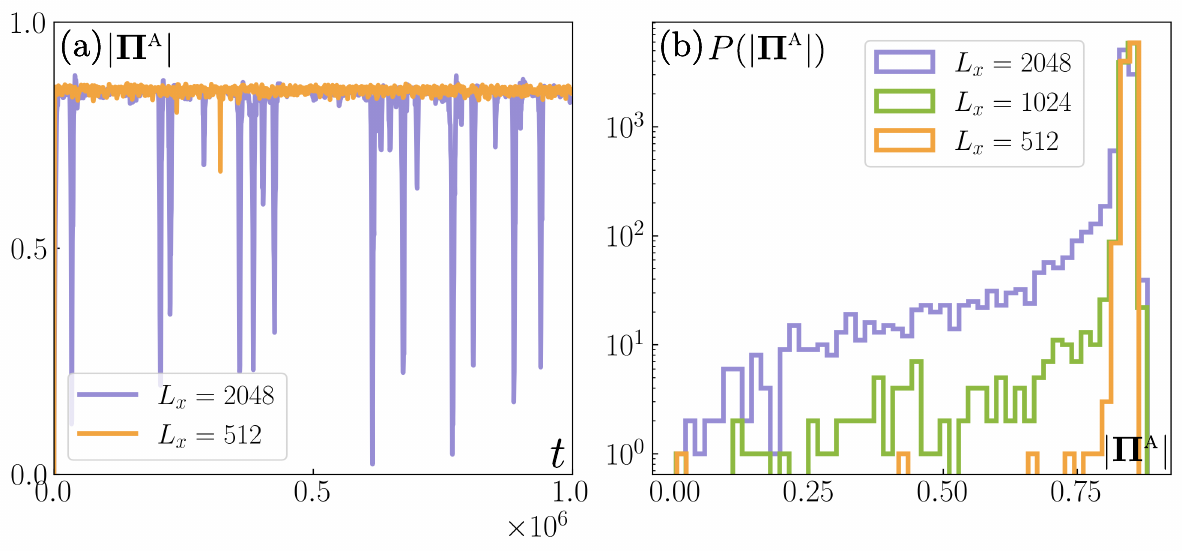}
    \caption{Instability of the laning pattern for mutually anti-aligning species.
    (a) Exemplary time series of the global polar order $|\bm \Pi^{\sa}|$ in the laning pattern for two system sizes.
    (b) The corresponding distributions.
    Parameters: $\eta = 0.30$ and $\Delta\chi = 0.35$.
    }
    \label{fig:figappA}
\end{figure}
%%%%%%%%%%%%%%%%%%%%%%%%%%%%%%%%%%

\setcounter{equation}{0}
\renewcommand{\theequation}{C\arabic{equation}}
\hypertarget{appC}{\textit{Appendix C: Derivation of the continuum model.---}} 
We coarse-grained a multi-population VM analogous to that defined in Eqs.~\eqref{micromodel} via the Boltzmann-Ginzburg-Landau (BGL) approach~\cite{bertinPRE2006,bertinjpa2009,Peshkov2014Jun}.
To keep the derivation general, we consider $n_{\sps}$ populations.
Particles of species $\sps$ self-propel at constant speed $v_0^{\sps}$ while we model the influence of angular noise via stochastic tumbling events happening at a frequency $\lambda^{\sps}$ and with a distribution $P_{\eta^{\sps}}(\theta)$ with zero mean and standard deviation $\eta^{\sps}$.

The BGL method consists in first writing a Boltzmann equation governing the evolution of the single-body distribution $f^{\sps}(\bm r,\theta,t)$ associated with species $\sps$.
Here, we have:
\begin{equation} \label{eq_Beq}
    \partial_t f^{\sps} + v_0^{\sps} \hat{\bm u}(\theta)\cdot \nabla f^{\sps} = 
    I_{\rm sd}^{\sps}[f^{\sps}] + I_{\rm al}^{\sps}[\{f^{\su}\}_{\su=1\ldots n_{\sps}}],
\end{equation}
where the $I_{\rm sd}^{\sps}$ and $I_{\rm coll}^{\sps}$ account for rotational self-diffusion and (anti-)alignment, respectively. These integrals read
\begin{align}
    & I_{\rm sd}^{\sps}[f^{\sps}] = \lambda^{\sps} \int_0^{2\pi} {\rm d}\phi \, f^{\sps}(\bm r,\phi,t)
    P_{\eta^{\sps}}(\theta - \phi)
    -\lambda^{\sps} f^{\sps}(\bm r,\theta,t),\\
%\end{align}
%\begin{align}
    & I_{\rm al}^{\sps}[\{f^{\su}\}]  = -\int_0^{2\pi}{\rm d}\phi\,
    \sum_{\su} f^{\sps}(\bm r,\theta,t)f^{\su}(\bm r,\phi,t)K^{\sps \su}(\theta - \phi) \nonumber \\
    & + \int_0^{2\pi}{\rm d}\phi_1\int_0^{2\pi}{\rm d}\phi_2 \,
    \sum_{\su} f^{\sps}(\bm r,\phi_1,t)f^{\su}(\bm r,\phi_2,t) \nonumber\\
    \label{eq_Icoll}
    & \times K^{\sps \su}(\phi_2 - \phi_1) P_{\eta^{\sps}}[\theta - \phi_1 - H^{\sps \su}(\phi_2 - \phi_1)].
\end{align}
Equation~\eqref{eq_Icoll} is written under the standard assumptions behind the formulation of the Boltzmann equation.
Namely, we neglect collisions involving more than two particles, while locality imposes that all distributions in~\eqref{eq_Icoll} depend on $\bm r$ only.
In addition, we used the molecular chaos hypothesis to factorize the two-body particle distribution.   

The kernel $K^{\sps\su}(\phi_2-\phi_1)$ is proportional to the rate of collisions between particles from species $\sps$ and $\su$ with pre-collisional orientations $\phi_1$ and $\phi_2$, respectively.
Following a similar approach as for single-species flocks~\cite{ChateLectureNotes} yields
%\begin{align*}
    $K^{\sps\su}(\phi_2-\phi_1) = 2r^{\sps}_0 \left|v_0^{\su}\hat{\bm u}(\phi_2) - v_0^{\sps}\hat{\bm u}(\phi_1)\right|$, 
    %\nonumber \\
    %& = 2 r^{\sps}_0 v_0^{\sps} \left[1 + \left(\frac{v_0^{\su}}{v_0^{\sps}}\right)^2 - 2\frac{v_0^{\su}}{v_0^{\sps}}\cos(\phi_2 - \phi_1) \right]^{\tfrac{1}{2}},
%\end{align*}
where $r_0^{\sps}$ is the radius of interaction associated with species $\sps$.
Furthermore, $\phi_1 + H^{\sps \su}(\phi_2 - \phi_1)$ corresponds to the post-collisional angle taken by both particles, such that we have
$H^{\sps \su}(\phi_2 - \phi_1) = \,{\rm arg}[ 1 + \chi^{\sps \su}e^{i(\phi_2-\phi_1)}]$.

Expanding the distribution $f^\sps$ into angular Fourier modes as $f^{\sps}(\bm r,\theta,t) = \frac{1}{2\pi}\sum_k f_k^{\sps}(\bm r,t) e^{-ik\theta}$, the latter obey [from Eq.~\eqref{eq_Beq}]
\begin{align}
     \label{eq_Beq_Fourier}
    \partial_t f^{\sps}_k + \frac{v_0^{\sps}}{2}\left[ \triangledown^* f_{k+1}^{\sps} + \triangledown f_{k-1}^{\sps}\right] =
    \tilde{\lambda}^{\sps}_k%(P_k^{\sps} - 1)
    f_k^{\sps}
    + \sum_{\su, q} J_{kq}^{\sps \su} f_q^{\su}f_{k-q}^{\sps},
\end{align}
where the complex gradient is defined as $\triangledown \equiv \partial_x + i\partial_y$, 
while $\tilde{\lambda}^{\sps}_k \equiv \lambda^{\sps}(P_k^{\sps} - 1)$
with $P_k^{\sps}$ the $k^{\rm th}$ Fourier mode of the distribution $P_{\eta^{\sps}}$. 
The mode coupling coefficients are given by $J_{kq}^{\sps \su} = P_k^{\sps} I_{kq}^{\sps \su} - I_{0 q}^{\sps \su}$ with
\begin{equation} \label{app_intI}
    I_{kq}^{\sps \su} \equiv \frac{1}{2\pi}\int_0^{2\pi}{\rm d}\phi\,
    K^{\sps \su}(\phi) e^{-i q \phi + i k H^{\sps \su}(\phi)}.
\end{equation}
The zeroth and first modes $f_0^{\sps}$ and $f_1^{\sps}$ are of particular interest, as they correspond to the complex representations of the local particle densities $\rho^{\sps} = f_0^{\sps}$ and polarities $\bm p^{\sps} = ({\rm Re}f_1^{\sps} \;\; {\rm Im}f_1^{\sps})^T$, respectively.
Equation~\eqref{eq_Beq_Fourier} defines an infinite hierarchy, which we close by assuming a standard scaling ansatz~\cite{bertinPRE2006,Peshkov2014Jun} describing the relative amplitudes of the Fourier modes and their space and time derivatives at the onset of global order.
Namely,
\begin{equation*} \label{app_ansatz}
    |f_k^{\sps}| \simeq \varepsilon^k \; (k>0), \qquad
    |f_0^{\sps} - \rho_0^{\sps}| \simeq \varepsilon, \qquad
    \partial_t \simeq \triangledown \simeq \varepsilon,
\end{equation*}
where $\varepsilon$ is a small parameter. 
Formally, this scaling ansatz supposes that all species exhibit comparable global polarities. This should remain true in the presence of weak population heterogeneity, which is the regime of interest here.

We truncate the hierarchy at the first nontrivial order $\varepsilon^3$.
The equations for the particle densities $f_0^{\sps} = \rho^{\sps}$ are then exact, 
while $f_1^{\sps}$ obeys
\begin{align}
     \partial_t f_1^{\sps} = & -\frac{v_0^{\sps}}{2}\triangledown\rho^{\sps} 
     -\frac{v_0^{\sps}}{2}\triangledown^* f_2^{\sps} 
     + \sum_{\su} M_1^{\sps\su}(\{\rho^{\sv}\}) f_1^{\su} \nonumber \\
     \label{eq_f1_complex_1}
     & + \sum_{\su} \left[ J_{12}^{\sps \su} f_2^{\su} f_1^{\sps*} + J_{1-1}^{\sps \su} f_2^{\sps} f_1^{\su*} \right] + O(\varepsilon^5),
\end{align}
where 
$M_k^{\sps\su}(\{\rho^{\sv}\}) \equiv [\tilde{\lambda}^{\sps}_k 
+ (\sum_{\sv} J_{k0}^{\sps \sv}\rho^{\sv})]\delta^{\sps\su} + J_{kk}^{\sps \su} \rho^{\sps}$.
The modes $f_2^{\sps}$ appear in Eq.~\eqref{eq_f1_complex_1} only through $O(\varepsilon^3)$ terms.
Hence, we write the equation for $f_2^{\sps}$ at $O(\varepsilon^2)$:
\begin{equation*}
    \label{app_eqf2}
    \sum_{\su} M_2^{\sps \su}(\{\rho_0^{\sv}\}) f_2^{\su} -\frac{v_0^{\sps}}{2}\triangledown f_1^{\sps} + \sum_{\su} J_{21}^{\sps \su} f_1^{\sps} f_1^{\su} + O(\varepsilon^3) = 0.
\end{equation*}
We close the hierarchy by solving this linear system for $f_2^{\sps}$ and inserting the solution into Eq.~\eqref{eq_f1_complex_1}.
Considering two species with identical speed $v_0$, interaction radius $r_0$ and noise strength $\eta$, 
we then recover Eqs.~\eqref{hydro} of the main text after adopting the vector representation $f_1^{\sps} \to \bm p^{\sps}$. 
In all numerical calculations, we used a Gaussian noise distribution such that $P_k^{\sps} = \exp[-k^2 \eta^2/2]$.

For arbitrary alignment couplings $\chi$, the integrals~\eqref{app_intI} do not admit simple analytical expressions, such that the coefficients in Eqs.~\eqref{hydro} cannot be simply expressed in terms of the parameters of the microscopic model. 
For convenience, we thus provide in~\cite{supplement} a Mathematica notebook that can be used to numerically evaluate the coefficients of the continuum model.

\setcounter{equation}{0}
\renewcommand{\theequation}{D\arabic{equation}}
\hypertarget{appD}{\textit{Appendix D: Enslaving of the non-hydrodynamic modes.---}} Here, we detail the enslaving procedure leading to Eq.~\eqref{mat_ensl}.
We consider a $m$-dimensional linear system of the form $\partial_t \bm X = \mathbb{M}(\bm q) \cdot \bm X$, where $\mathbb{M}(\bm q)$ is a wave-vector dependent $m\times m$ matrix. 
As we are interested in the $q = |\bm q| \to 0$ limit, we expand $\mathbb{M}$ 
as 
\begin{equation} \label{app_Mexp}
    \mathbb{M}(\bm q) = \mathbb{M}_0 + i q \mathbb{M}_1 - q^2 \mathbb{M}_2 + \mathcal{O}(q^3),
\end{equation}
where the $\mathbb{M}_i$ matrices are independent of $q$ but may depend on the direction of $\bm q$.

We denote $\{\bm Y_{j}\}$ as the left eigenbasis of $\mathbb{M}_0$, such that for all $j = 1, \ldots, m$, $\bm Y_j^T \cdot \mathbb{M}_0 = \Lambda_j \bm Y_j^T$,
where $\{\Lambda_j\}$ are the corresponding eigenvalues.
Assuming that the initial linear system describes fluctuations around a solution which is stable under $q = 0$ perturbations, all $\Lambda_j \le 0$.
We now suppose that the problem has $n_{\rm h}$ hydrodynamic modes, and we reorder the eigenvalues such that
$\Lambda_1, \ldots, \Lambda_{n_{\rm h}} = 0$ and $\Lambda_{n_{\rm h}+1}, \ldots, \Lambda_{m} < 0$.
Defining 
\begin{equation} \label{app_tildeX}
    \Tilde{\bm X} = \begin{pmatrix} \bm Y_1 & \cdots & \bm Y_m\end{pmatrix}^T \cdot \bm X \equiv \mathbb{P} \cdot \bm X,
\end{equation}
the first $n_{\rm h}$ elements of $\Tilde{\bm X}$ are the hydrodynamic modes, while the other $m -n_{\rm h}$ are fast.
This is straightforwardly shown from
\begin{align*}
    \partial_t \Tilde{\bm X} & \underset {q = 0}{=} \mathbb{P} \cdot \partial_t \bm X = \mathbb{P} \cdot \mathbb{M}_0 \cdot \bm X 
    %= {\rm diag}(\lambda_1 \dots \lambda_m)\cdot \mathbb{P} \cdot \bm X \\
    %& 
    = {\rm diag}(\lambda_1 \dots \lambda_m) \cdot \Tilde{\bm X},
\end{align*}
where we have used that $\mathbb{P} \cdot \mathbb{M}_0 \cdot \mathbb{P}^{-1} = {\rm diag}(\lambda_1 \dots \lambda_m)$.
While the slow modes are uniquely defined, the fast modes may be expressed differently so long as they are associated with a nonzero damping rate. The formulation~\eqref{app_tildeX} nevertheless provides a compact change of basis formula between the slow and fast modes and the original variables.
Going back to $q > 0$, the dynamics of $\Tilde{\bm X}$ then follows
$\partial_t\Tilde{\bm X} = \tilde{\mathbb{M}}(\bm q)\cdot \Tilde{\bm X}$, where $\Tilde{\mathbb{M}} \equiv \mathbb{P}\cdot \mathbb{M} \cdot \mathbb{P}^{-1}$. 
In the regime of interest, $\mathbb{P}$ is always invertible. 
For convenience, we write separately the dynamics of the hydrodynamic ($\bm X_{\rm h}$) and fast ($\bm X_{\rm f}$) modes (dropping the tildes to lighten notations):
\begin{align}
        \partial_t\bm X_{\rm h} - \mathbb{M}^{\rm h}_{\rm h} \cdot \bm X_{\rm h} 
        & = \mathbb{M}^{\rm h}_{\rm f} \cdot \bm X_{\rm f}  ,\label{app_evol_slow_modes}\\
        \partial_t\bm X_{\rm f}  - \mathbb{M}^{\rm f} _{\rm f}  \cdot \bm X_{\rm f} 
        & = \mathbb{M}^{\rm f}_{\rm h} \cdot \bm X_{\rm h},\label{app_evol_fast_modes}
\end{align}
in terms of the block matrices $\mathbb{M}^x_y$.
By construction, all elements of $\mathbb{M}^{\rm h}_{\rm h}$ and $\mathbb{M}^{\rm h}_{\rm f}$ are zero for $q = 0$, while $\mathbb{M}^{\rm f}_{\rm f}(q=0)$ is diagonal with negative entries representing the finite damping rates of the fast modes.
Expanding all block matrices similarly to~\eqref{app_Mexp}, we then enslave the fast modes by solving Eq.~\eqref{app_evol_fast_modes} in the limit $q \to 0$.
Here, we focus on the leading order solution, while a derivation including ${\cal O}(q^2)$ corrections is presented in~\cite{supplement}.
Therefore,
\begin{align*}
    \bm X_{\rm f} = -\left({\mathbb{M}}^{\rm f}_{{\rm f},0}\right)^{-1}\cdot {\mathbb{M}}^{\rm f}_{{\rm h},0} \cdot \bm X_{\rm h} + \mathcal{O}(q).
\end{align*}
Replacing this solution in Eq.~\eqref{app_evol_slow_modes},
we then find that the hydrodynamic modes obey $\partial_t\bm X_{\rm h} = \mathbb{M}_{\rm h}(\bm q) \cdot \bm X_{\rm h}$ with
\begin{equation}
    \mathbb{M}_{\rm h}(\bm q) = iq\left( \mathbb{M}^{\rm h}_{{\rm h},1} - \mathbb{M}^{\rm h}_{{\rm f},1}\cdot \left({\mathbb{M}}^{\rm f}_{{\rm f},0}\right)^{-1}\cdot {\mathbb{M}}^{\rm f}_{{\rm h},0} \right) + {\cal O}(q^2).
\end{equation}

Performing this analysis for the linear stability described in the main text, we find three slow modes: $\delta\hat{\rho}^{\sa,\spb}$ and $\bar{\theta} = s^\sa_0 F^{\spb}_{\sa}\delta \hat{\theta}^\sa + {s^\spb_0}{F^{\sa}_{\spb}}\delta\hat{\theta}^\spb$.
The expressions of the various matrices involved in the derivation are quite lengthy and not very informative.
We can nevertheless express the coefficients that couple the density fluctuations in Eq.~\eqref{mat_ensl} as
\begin{align*}
    V_{\sa\sa} & = v_0^{\sa} \left[ C_{\sa\sa} E^{\sa}_{\sa} + C_{\sa\spb} E^{\spb}_{\sa} \right], &
    V_{\sa\spb} & = v_0^{\sa} \left[ C_{\sa\sa} E^{\sa}_{\spb} + C_{\sa\spb} E^{\spb}_{\spb} \right],\\
    V_{\spb\sa} & = v_0^{\spb} \left[ C_{\spb\sa} E^{\sa}_{\sa} + C_{\spb\spb} E^{\spb}_{\sa} \right], &
    V_{\spb\spb} & = v_0^{\spb} \left[ C_{\spb\sa} E^{\sa}_{\spb} + C_{\spb\spb} E^{\spb}_{\spb} \right],
\end{align*}
where the $C_{\sps\su}$ coefficients are constant and depend on the model parameters, while $E^{\sps}_{\su} = \left(s_0^{\sa}\partial_{\rho^{\su}}\alpha^{\sps}_{\sa} + \epsilon s_0^{\spb}\partial_{\rho^{\su}}\alpha^{\sps}_{\spb}\right)_{\rho^{\sps} = \rho_0}$ with $\epsilon = {\rm sign}(\bar{\chi})$~\cite{supplement}.

\clearpage

\onecolumngrid 
\normalfont
\renewcommand \thefigure{S\arabic{figure}}
\setcounter{figure}{0}
\setcounter{equation}{0}
\renewcommand{\theequation}{S\arabic{equation}}
\begin{center}
\textbf{\Large Supplemental Material: Nonreciprocity as a Generic Mechanism for Demixing in Flocking Mixtures}
\end{center}
\vspace{2ex}
\input{supplement_arxiv.tex}

\end{document}

%% file: supplement_arxiv.tex
In this document, we provide details on the construction of the phase diagrams shown in Fig. 1 of the main text, together with additional representative snapshots. We also expand on the continuum theory; specifically, we present exact expressions for the hydrodynamic equation coefficients, the complete linear stability analysis, and details on the enslaving of the non-hydrodynamic modes. Finally, we include a description of the accompanying supplemental movies.

\section{Agent-based simulations}

\subsection{Phase diagrams}
\begin{figure}[h!]
    \centering
    \includegraphics[width=0.8\linewidth]{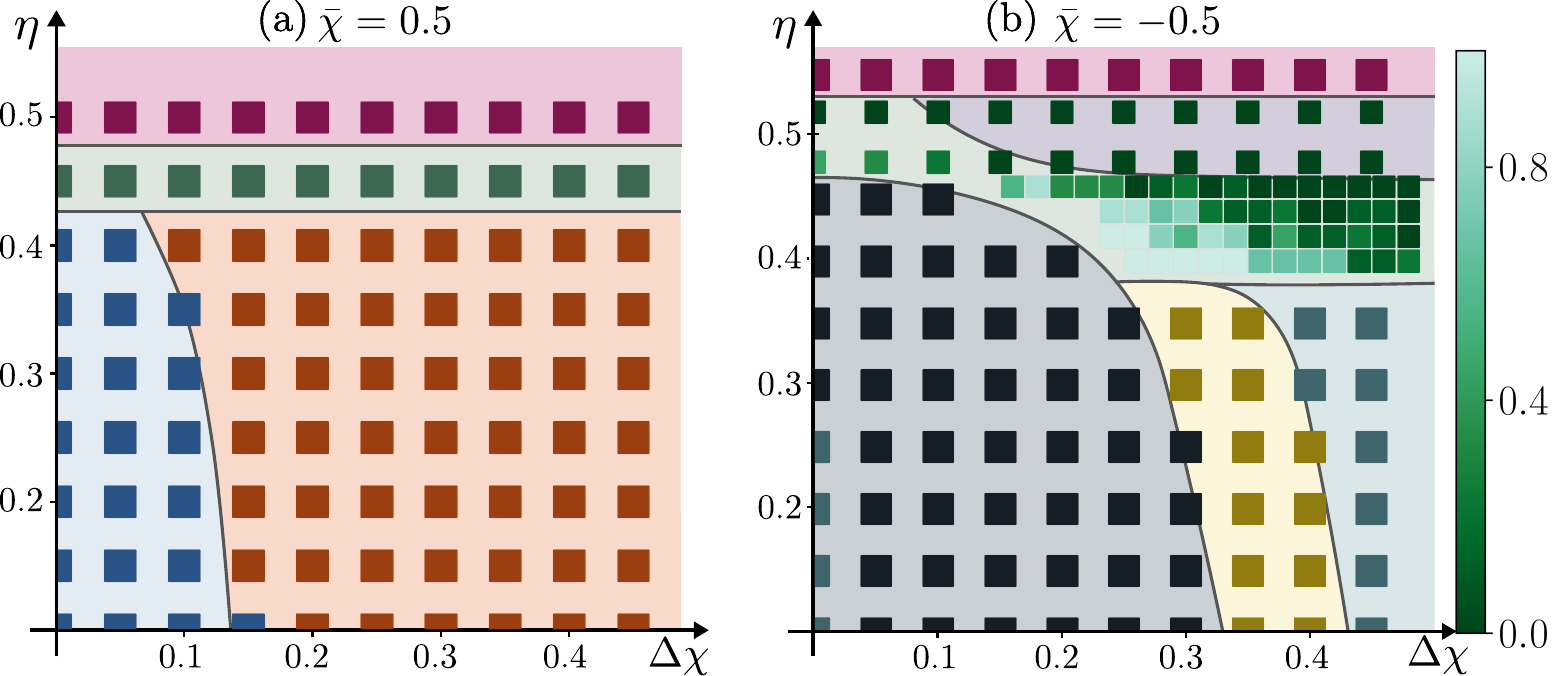}
    \caption{Finite-size phase diagrams in the 
 $(\Delta\chi,\eta)$ plane for aligning (a) and anti-aligning (b) interspecies interactions. Each square corresponds to one simulation, color-coded by the identified steady-state phase.
 On the left: disordered (pink), homogeneous polar (blue), Vicsek bands (green) and demixed single band (orange).
 On the right: disordered (pink), homogeneous apolar (grey), polar and apolar band patterns (green), laning (yellow), and chaotic clusters (blue).
 For $\bar\chi = -\tfrac{1}{2}$, shades of green indicate the fraction of apolar band configurations observed among 10 independent simulations (scale on the right side of the figure).}
    \label{fig:figS1}
\end{figure}

The stylized finite-size phase diagrams in Fig. 1(a,b) of the main text were constructed by systematically scanning the $(\Delta\chi,\eta)$ space. Every square in Fig. \ref{fig:figS1} corresponds to an independent simulation.
The corresponding phase, marked by a single color, was assigned based on visual inspection of multiple steady-state snapshots. Phase boundaries were determined by separating points of different colors. As discussed in the main text, for mutually anti-aligning populations we find, near the onset of order, coexistence between polar configurations with bands of alternating species moving in the same direction (polar band pattern) and apolar configurations with bands traveling in opposite directions (apolar band pattern). In this region, 10 independent simulations were performed starting from isotropic initial conditions, and the color of each square denotes the fraction of runs displaying an apolar band pattern. 

\begin{figure}
    \centering
    \includegraphics[width=0.75\linewidth]{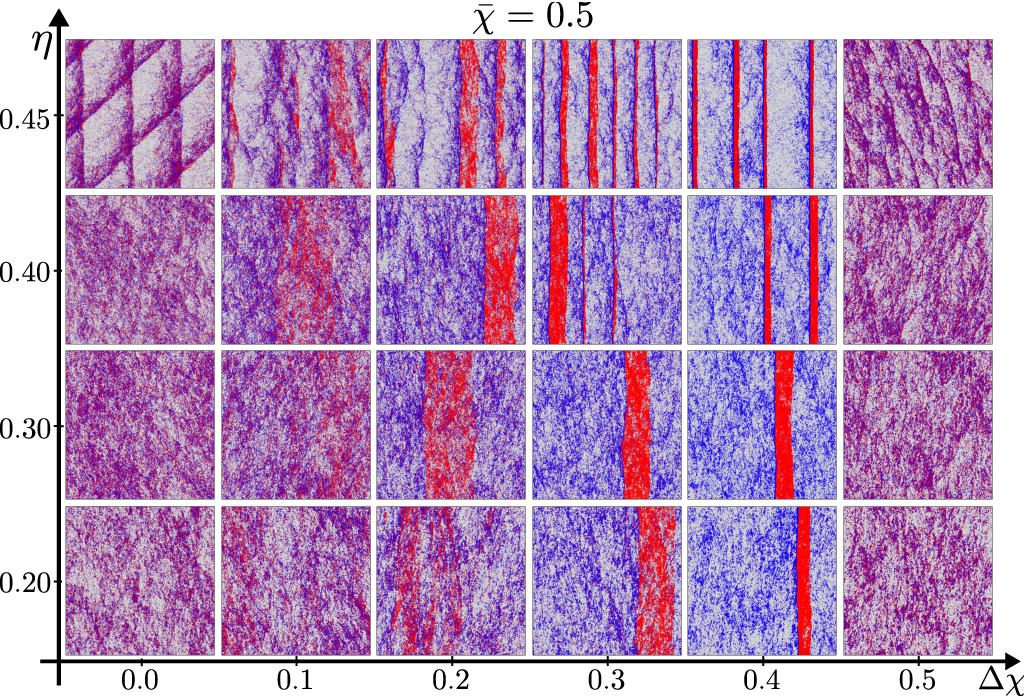}
    \caption{Representative snapshots in the $(\Delta\chi,\eta)$ plane for mutually aligning intraspecies interactions. Simulation parameters are $\rho_0 = \frac{1}{2}$, $v_0 = 1, \bar\chi = \frac{1}{2}$ and $L=512$. Particles are color-coded by species with $\sa$ in red and $\spb$ in blue.}
    \label{fig:figS2}
\end{figure}

\begin{figure}
    \centering
    \includegraphics[width=0.82\linewidth]{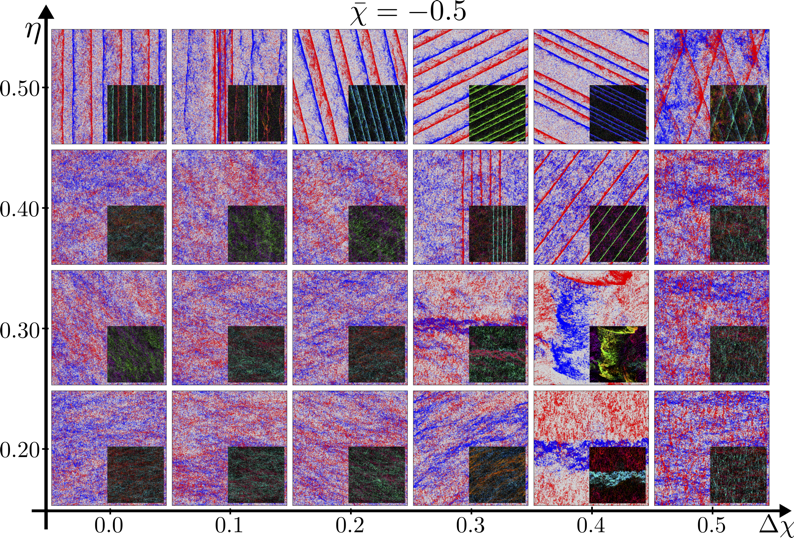}
    \caption{Representative snapshots in the $(\Delta\chi,\eta)$ plane for mutually anti-aligning intraspecies interactions. Simulation parameters are $\rho_0 = \frac{1}{2}$, $v_0 = 1, \bar\chi = -\frac{1}{2}$ and $L=512$. Particles are color-coded by species in the main panels ($\sa$ in red and $\spb$ in blue), and by orientation in the insets.}
    \label{fig:figS3}
\end{figure}

The parameters of the representative snapshots in Fig. 1(e-f) of the main text are $\rho_0 = \frac{1}{2}$, $v_0 = 1$ and $\bar\chi = \pm\frac{1}{2}$, with (c) $\eta = 0.45$, $\Delta\chi = 0.45$, (d) $\eta = 0.30$, $\Delta\chi = 0.40 $, (e) $\eta = 0.45$, $\Delta\chi = 0.30$, (f) $\eta = 0.50$, $\Delta\chi = 0.45$, (g) $\eta = 0.30$, $\Delta\chi = 0.45$ and (h) $\eta = 0.25$, $\Delta\chi = 0.35$. Additional representative snapshots are shown in Figures \ref{fig:figS2} and \ref{fig:figS3}, for mutually aligning and anti-aligning interspecies interactions, respectively. 

\subsection{Phase coexistence bordering the transition for mutually aligning species}

\begin{figure}[!h]
    \centering
    \includegraphics[width=0.65\linewidth]{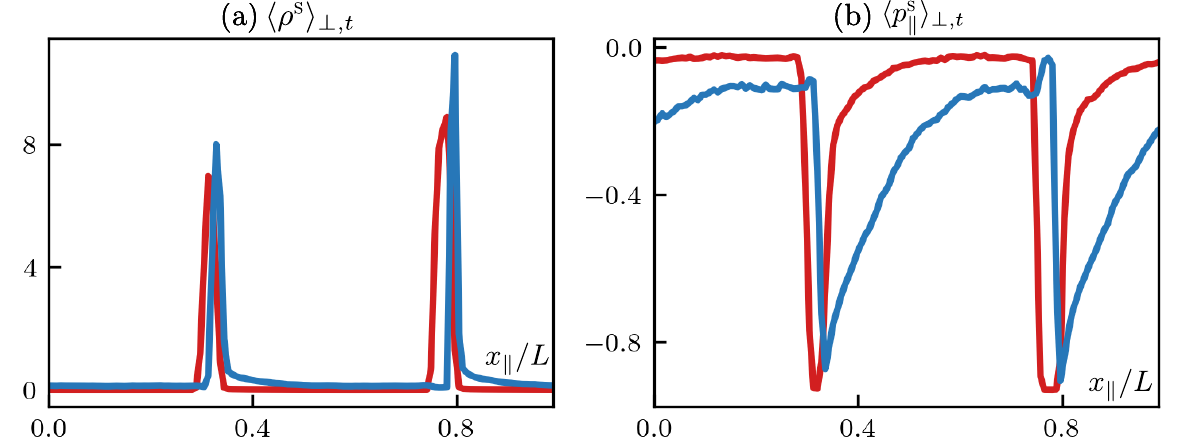}
    \caption{Characterization of the bands phase, which emerges for $\bar\chi > 0$ and high noise $\eta$. (a) Density profiles $\rho^\sps$ and (b) longitudinal component of the polarization $\bm p^\sps$, both averaged along the band and over time. Parameters: $\eta = 0.45, \Delta\chi = 0.45$ and $L = 512$.  }
    \label{fig:figS4}
\end{figure}

Here, we provide additional details on the band phase for mutually aligning species (panels with $\eta= 0.45$ in Fig.~\ref{fig:figS2}). In this region, reminiscent of the bands observed in the single-species Vicsek model, dense bands propagate through a dilute, gaseous background. For two species with reciprocal interactions, these bands generically contain a mixture of both species. Increasing the nonreciprocity, however, induces a separation between red and blue bands. Each structure is then composed of a leading band primarily formed by the species with the stronger interspecies coupling (red in Fig.~\ref{fig:figS4}), followed closely by a band dominated by the other species (blue in Fig.~\ref{fig:figS4}). Note that the panels in Fig.~\ref{fig:figS2} with $\eta = 0.40$ and $\Delta\chi = 0.3, 0.4$ correspond to the single-band phase, as evidenced by the red band traveling through a homogeneous blue liquid, although the coarsening dynamics are slow in this regime.

\section{Continuum theory}

\subsection{Coefficients}

\begin{figure}
    \centering
    \includegraphics[width = 0.7\linewidth]{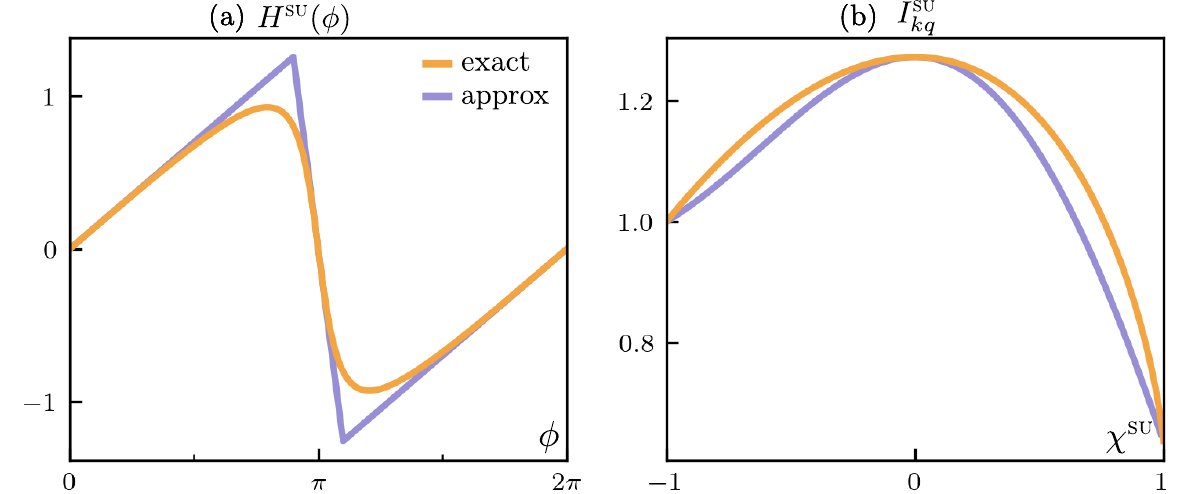}
    \caption{Comparison between the exact and approximated expressions of (a) the interaction kernel $H^{\sps\su}(\phi)$ for $\chi^{\sps\su} = 0.8$, and (b) the integral $I^{\sps\su}_{kq}$ as a function of $\chi^{\sps\su}$, with $k=1$ and $q=0$.}
    \label{fig:figS5}
\end{figure}

As mentioned in the main text, for arbitrary alignment couplings $\chi$, the integrals 
\begin{equation} \label{sm_intI}
    I_{kq}^{\sps \su} \equiv \frac{1}{2\pi}\int_0^{2\pi}{\rm d}\phi\,
    K^{\sps \su}(\phi) e^{-i q \phi + i k H^{\sps \su}(\phi)}.
\end{equation}
do not admit simple analytical expressions. This is because, in our multi-species model (in contrast to the single-species Vicsek model), the interaction kernel $H^{\sps \su}(\phi_2 - \phi_1) = \,{\rm arg}[ 1 + \chi^{\sps \su}e^{i(\phi_2-\phi_1)}]$ is not a linear function of $\phi$ (see Figure \ref{fig:figS5}(a)). However, $H^{\sps\su}(\phi)$ can be approximated by a piecewise linear function:
\begin{equation}
    H^{\sps\su}_{\textrm{approx}}(\phi) = \begin{cases} 
      \frac{\chi^{\sps\su}}{1+\chi^{\sps\su}}\phi & 0 \leqslant \phi < \frac{\pi}{2}(1+\chi^{\sps\su}) \\
      -\frac{\chi^{\sps\su}}{1-\chi^{\sps\su}}(\phi -\pi) & \frac{\pi}{2}(1+\chi^{\sps\su}) \leqslant \phi \leqslant \frac{\pi}{2}(3 - \chi^{\sps\su}) \\
      \frac{\chi^{\sps\su}}{1+\chi^{\sps\su}}(\phi - 2\pi) & \frac{\pi}{2}(3 -\chi^{\sps\su}) < \phi \leqslant 2\pi .\\
   \end{cases}
\end{equation}
Using this approximation of the kernel, the integrals~\eqref{sm_intI} become
\begin{align*}
I_{kq, \textrm{approx}}^{\sps\su} &= \frac{1}{\pi}\int_{0}^{\frac{\pi}{2}(1
+ \chi^{\sps\su})}\sin\left(\frac{\phi}{2}\right)\exp\left(i\left(-q+k\frac{\chi^{\sps\su}}{1+\chi^{\sps\su}}\right)\phi\right){\rm d}\phi \\ 
    & + \frac{1}{\pi}\int_{\frac{\pi}{2}(1
+ \chi^{\sps\su})}^{\frac{\pi}{2}(3
- \chi^{\sps\su})}\left|\sin\left(\frac{\phi}{2}\right)\right|\exp\left(-i\left(q\phi+k\frac{\chi^{\sps\su}}{1-\chi^{\sps\su}}(\phi- \pi)\right)\right){\rm d}\phi \\ 
    &+ \frac{1}{\pi}\int_{\frac{\pi}{2}(3
- \chi^{\sps\su})}^{2\pi}\sin\left(\frac{\phi}{2}\right)\exp\left(i\left(-q\phi+k\frac{\chi^{\sps\su}}{1+\chi^{\sps\su}}(\phi- 2\pi)\right)\right){\rm d}\phi 
\end{align*}
 Figure~\ref{fig:figS5}(b) shows the exact and approximated values of the integral as functions of $\chi^{\sps\su}$ for $k= 1$ and $q = 0$. This approximation greatly accelerates the numerical evaluation of the integrals, and we have verified that it does not lead to any quantitative change in our results. 
 
In terms of these integrals, the coefficients appearing in Eqs.~(2) of the main text are given by  
\begin{align*}
    & \psi_{\su\sv}^{\sps} = \kappa_{1,\su\sv}^{\sps} + \kappa_{1,\sv\su}^{\sps} + \kappa_{2,\sv\su}^{\sps}, &
    & \lambda_{\su\sv}^{\sps} = \kappa_{1,\su\sv}^{\sps} + \kappa_{1,\sv\su}^{\sps} - \kappa_{2,\su\sv}^{\sps}, \\
    & \nu_{\su\sv}^{\sps} = 
    \kappa_{2,\su\sv}^{\sps} -
    \kappa_{1,\su\sv}^{\sps} -\kappa_{1,\sv\su}^{\sps}, &
    & \xi^{\sps}_{\su\sv\sw} = 
    \frac{1}{2}\left[ \beta^{\sps}_{\sv\su\sw}
    + \beta^{\sps}_{\sw\su\sv}
    + \beta^{\sps}_{\sw\sv\su}
    + \beta^{\sps}_{\sv\sw\su}
    - \beta^{\sps}_{\su\sv\sw}
    - \beta^{\sps}_{\su\sw\sv}
    \right], \\
    &D^{\sps}_{\su} = -\frac{v_0^{\sps}}{2}\gamma^{\sps}_{\su}, &
    & \alpha^{\sps}_{\su} = \left[ \tilde{\lambda}^{\sps}_1 + \sum_{\spt} J_{10}^{\sps\spt} \rho^{\spt} \right]\delta_{\sps\su} +  J_{11}^{\sps\su} \rho^{\sps}, 
\end{align*}
where 
\begin{align*}
    & J_{kq}^{\sps \su} = P_k^{\sps} I_{kq}^{\sps \su} - I_{0 q}^{\sps \su}, & 
    & \tilde{\lambda}^{\sps}_k \equiv \lambda^{\sps}(P_k^{\sps} - 1)\\
    & \beta^{\sps}_{\su \sv \sw}  = - J_{1-1,0}^{\sps \su}\zeta^{\sps}_{\sv \sw} - \delta_{\sps\su}\sum_{\spt} J_{12,0}^{\sps \spt}\zeta^{\spt}_{\sv \sw}, & 
    & \kappa_{2, \su\sv}^{\sps}  = -J_{1-1,0}^{\sps \su}\gamma^{\sps}_{\sv} - \delta_{\sps \su}\sum_{\spt} J_{12,0}^{\sps \spt}\gamma^{\spt}_{\sv},\\
    &\kappa_{1,\su\sv}^{\sps}  =\frac{v_0^{\sps}}{2}\zeta^{\sps}_{\su\sv}, &
    &\zeta^\sps_{\su\sv} = -M_2^{-1}[\sps,\su]J^{\su\sv}_{21}\\
    & \gamma^{\sps}_{\su} = M_2^{-1}[\sps,\su]\frac{v_0^\su}{2}  & 
    &  M_k^{\sps\su} = [\tilde{\lambda}^{\sps}_k 
+ (\sum_{\sv} J_{k0}^{\sps \sv}\rho^{\sv})]\delta^{\sps\su} + J_{kk}^{\sps \su} \rho^{\sps}.
\end{align*}
Here, $M_2^{-1}[\sps,\su]$ means the component $\sps\su$ of the inverted matrix $M_2$. For convenience, we also provide a Mathematica notebook that can be used to numerically evaluate these coefficients.

\subsection{Hydrodynamic equations}
As mentioned in the main text, it is convenient to express the polarities as $\bm p^\sps = s^\sps \hat{\bm u}(\theta^\sps)$ and to rewrite Eqs. (2) in terms of their magnitudes $s^\sps$ and orientations $\theta^\sps$. In compact form, the resulting equations can be written as
\begin{subequations}
\label{eq_s_theta_nspecies_compact}
\begin{align} 
    \partial_t s^{\sps}
    & + \sum_{\su} \left[ \bm K_\su^\sps \cdot \nabla s^\su 
    + s^\su \bm L_\su^\sps \cdot \nabla\theta^\su \right]
     = -\frac{v_0^{\sps}}{2}(\hat{\bm u}^\sps\cdot\nabla) \rho^{\sps}
    + \sum_{\su}
    F^\sps_\su\cos(\Delta\theta_{\sps\su})s^{\su} \nonumber \\
    & + \sum_\su D^{\sps}_{\su} \left[ 
    \cos(\Delta\theta_{\sps\su})\left( 
    \Delta s^\su - s^\su |\nabla\theta^\su|^2
    \right)
    + \sin(\Delta\theta_{\sps\su})\left( 
    2(\nabla s^\su)\cdot(\nabla \theta^\su) + s^\su \Delta\theta^\su
    \right)
    \right], \\
    s^{\sps} \partial_t \theta^{\sps}
    & + \sum_{\su} \left[ -\bm L_\su^\sps \cdot \nabla s^\su 
    + s^\su \bm K_\su^\sps \cdot \nabla\theta^\su \right]
     = -\frac{v_0^{\sps}}{2}(\hat{\bm u}_\perp^\sps\cdot\nabla) \rho^{\sps}
    - \sum_{\su}
    F^\sps_\su\sin(\Delta\theta_{\sps\su})s^{\su} \nonumber \\
    & + \sum_\su D^{\sps}_{\su} \left[ 
    -\sin(\Delta\theta_{\sps\su})\left( 
    \Delta s^\su - s^\su |\nabla\theta^\su|^2
    \right)
    + \cos(\Delta\theta_{\sps\su})\left( 
    2(\nabla s^\su)\cdot(\nabla \theta^\su) + s^\su \Delta\theta^\su
    \right)
    \right],
\end{align}
\end{subequations}
 where $F^{\sps}_{\su} \equiv \alpha^{\sps}_{\su}(\rho_0^{\sa},\rho_0^{\spb}) - 
    \sum_{\sv,\sw}\xi^{\sps}_{\su\sv\sw} s_0^\sv s_0^\sw \cos(\Delta\theta_{\sv\sw})$,  $\Delta\theta_{\su\sv} \equiv \theta^\su - \theta^\sv$,  $\hat{\bm u}_\perp^\sps = (-\sin(\theta^\sps), \cos(\theta^\sps))^T$ and
\begin{align*}
    \bm K_\su^\sps & \equiv 
    \cos(\Delta\theta_{\sps\su}) \sum_\sv \psi_{\su\sv}^{\sps} s^\sv \hat{\bm u}^\sv
    + \hat{\bm u}^\su \sum_\sv \lambda_{\sv\su}^{\sps} 
    s^\sv \cos(\Delta\theta_{\sps\sv})
    + \hat{\bm u}^\sps \sum_\sv \nu_{\sv\su}^{\sps} 
    s^\sv \cos(\Delta\theta_{\su\sv}), \\
    \bm L_\su^\sps & \equiv 
    \sin(\Delta\theta_{\sps\su}) \sum_\sv \psi_{\su\sv}^{\sps} s^\sv \hat{\bm u}^\sv
    + \hat{\bm u}_\perp^\su \sum_\sv \lambda_{\sv\su}^{\sps} 
    s^\sv \cos(\Delta\theta_{\sps\sv})
    - \hat{\bm u}^\sps \sum_\sv \nu_{\sv\su}^{\sps} 
    s^\sv \sin(\Delta\theta_{\su\sv}).\\
\end{align*}
\subsection{Homogeneous stationary solutions}

For two species ($\sa$ and $\spb$), the homogeneous stationary solutions of these equations have a constant density $\rho^{\sps} = \rho^{\sps}_0$ and $s^{\sps} = s_0^{\sps}$, and solve  for $\su \ne \sps$
\begin{align}
    \label{sm_homsol}
    F^{\sps}_{\sps}s_0^\sps + F^{\sps}_{\su}s_0^\su \cos(\Delta\theta_{\sps\su}) = 0 , 
    \quad
    F^{\sps}_{\su}s_0^{\su}\sin(\Delta\theta_{\sps\su}) = 0,
\end{align}
In the regime where $\chi^{\sa\spb}$ and $\chi^{\spb\sa}$ have the same sign, the equations admit 3 types of steady-state solutions. For sufficiently high noise (with the threshold depending on the density), the only solution is the disorder state for which $s^{\sa,\spb} = 0$. As  the noise decreases, the disordered solution becomes unstable, and the solutions bifurcate to an ordered state. Depending on the alignment couplings, two ordered phases appear, a flocking state with $s^{\sa,\spb} = s^{\sa,\spb}_0$ and $\Delta\theta_{\spb\sa} = 0$, and an anti-flocking state with $s^{\sa,\spb} = s^{\sa,\spb}_0$ and $\Delta\theta_{\spb\sa} = \pi$.
In fact, both solutions generally exist for small $\bar\chi$. However, the flocking and anti-flocking solutions are only stable to homogeneous perturbations for $\bar\chi > 0$ and $\bar\chi < 0$, respectively. For the linear stability analysis with spatially-dependent perturbations, we consider only these solutions. Since Eqs.~\eqref{sm_homsol} are nonlinear, we determine $s^{\sa,\spb}_0$ numerically.

\subsection{Linear stability analysis}

To analyse the stability of the homogeneous steady-state solutions, we examine the evolution of small linear perturbations around these states. Specifically, we consider perturbations $(\rho^\sa,\rho^\spb, s^\sa,s^\spb, \theta^\sa,\theta^\spb) = (\rho_0^\sa + \delta\rho^\sa ,\rho_0^\spb + \delta\rho^\spb,s^\sa_0 + \delta s^\sa,s^\spb_0 + \delta s^\spb,\theta^\sa_0 + \delta\theta^\sa,\theta^\spb_0+ \delta\theta^\spb)$ and only keep terms up to linear order. For convenience, we choose a reference frame such that $\theta^\sa_0 = 0$ and $\theta^\spb_0 = \Delta\theta_0$, and we express the linearized equations in terms of $\delta\theta^\sps \to s_0^\sps \delta\theta^\sps$. The linearized equations then read
\begin{subequations}
\label{sm_eq_fluc}
\label{eq_lin_system}
    \begin{align} \label{sm_eq_rho_fluc}
    & \partial_t \delta\rho^\sps + v_0^\sps \hat{\bm u}_0^\sps \cdot \nabla \delta s^\sps + v_0^\sps \hat{\bm u}_{\perp,0}^\sps \cdot \nabla \delta \theta^\sps = {\cal D}^\sps \nabla^2 \delta\rho^\sps, \\
    & \partial_t \delta s^{\sps}
     + \sum_{\su} \left[ \bm K_{\su,0}^\sps \cdot \nabla \delta s^\su 
    + \bm L_{\su,0}^\sps \cdot \nabla \delta \theta^\su \right]
     = -\frac{v_0^{\sps}}{2}(\hat{\bm u}_0^\sps\cdot\nabla) \delta\rho^{\sps} 
    + \sum_{\su} \left[ 
    F^\sps_{\su,0}\cos(\Delta\theta_{\sps\su,0})\delta s^{\su}
    - F^\sps_{\su,0}\sin(\Delta\theta_{\sps\su,0}) \left(\frac{s^{\su}_0}{s^{\sps}_0}\delta\theta^\sps - \delta\theta^\su\right) \right. \nonumber \\
    & \left. + \cos(\Delta\theta_{\sps\su,0})s^{\su}_0 \left( 
    \sum_\sv F^\sps_{\su,\rho^\sv} \delta\rho^\sv
    + F^\sps_{\su,s^\sv} \delta s^\sv
    + F^\sps_{\su,\theta^\sv} \frac{\delta\theta^\sv}{s^{\sv}_0}
    \right)\right]
    + \sum_\su D^{\sps}_{\su} \left[ 
    \cos(\Delta\theta_{\sps\su,0}) \Delta \delta s^\su
    + \sin(\Delta\theta_{\sps\su,0}) \Delta\delta\theta^\su
    \right], \\
    & \partial_t \delta\theta^{\sps}
    + \sum_{\su} \left[ -\bm L_{\su,0}^\sps \cdot \nabla \delta s^\su 
    + \bm K_{\su,0}^\sps \cdot \nabla \delta \theta^\su \right]
     = -\frac{v_0^{\sps}}{2}(\hat{\bm u}_{\perp,0}^\sps\cdot\nabla) \delta\rho^{\sps} \nonumber \\
    & - \sum_{\su} \left[F^\sps_{\su,0}\sin(\Delta\theta_{\sps\su,0})\delta s^{\su}
    + F^\sps_{\su,0}\cos(\Delta\theta_{\sps\su,0}) \left(\frac{s^{\su}_0}{s^{\sps}_0}\delta\theta^\sps - \delta\theta^\su\right)
    + \sin(\Delta\theta_{\sps\su,0})s^{\su}_0 \left( 
    \sum_\sv F^\sps_{\su,\rho^\sv} \delta\rho^\sv
    + F^\sps_{\su,s^\sv} \delta s^\sv
    + F^\sps_{\su,\theta^\sv} \frac{\delta\theta^\sv}{s_0^\sv}
    \right)\right] \nonumber \\
    & + \sum_\su D^{\sps}_{\su} \left[ 
    -\sin(\Delta\theta_{\sps\su,0}) \Delta \delta s^\su
    + \cos(\Delta\theta_{\sps\su,0}) \Delta\delta\theta^\su
    \right].
    \end{align}
\end{subequations}
Note that we have added a diffusion term on the r.h.s.\ of Eq.~\eqref{sm_eq_rho_fluc}. 
In the single species case, adding translational diffusion at the level of the Boltzmann equation ---which results in additional diffusion for both $\rho^\sps$ and $\bm p^\sps$---reduces the occurrence of spurious instabilities of the ordered phase~\cite{Peshkov2014Jun}.
For all numerical evaluations of the coefficients in Eqs.~\eqref{sm_eq_fluc}, we have added a diffusion term with coefficient $D = 0.5$, resulting in $ {\cal D}^\sa = {\cal D}^\spb = 0.5$, as well as 
$D^\sa_\sa \to D^\sa_\sa + 0.5$ and $D^\spb_\spb \to D^\spb_\spb + 0.5$.

In Eqs.~\eqref{sm_eq_fluc}, the subscript $0$ indicates parameters evaluated at the steady state solution, while we use the shorthand notation $F^\sps_{\su,\alpha} \equiv \partial_\alpha F^\sps_{\su}|_0$ with $\alpha = \rho^\sv$, $s^\sv$ or $\theta^\sv$. Going to Fourier space and recasting Eqs. \eqref{eq_lin_system} into a linear system, we find that it takes the form $\partial_t \hat{\bm X} = \mathbb{M}(q)\hat{\bm X}$ where $\hat{\bm X} \equiv \begin{pmatrix} \delta \hat\rho^\sa & \delta\hat\rho^\spb & \delta\hat\theta^\sa & \delta\hat\theta^\spb &
    \delta \hat{s}^\sa & 
    \delta \hat{s}^\spb \end{pmatrix}^T$ and the rows of the six-dimensional matrix $\mathbb{M}(\bm q)$ are given by

\begin{align*}
    & \mathbb{M}^{\rho^\sa}= \begin{pmatrix}
        - \mathcal{D}^\sa q^2 & 0 & i v_0^\sa q_x & 0 & i v_0^\sa q_y & 0 
    \end{pmatrix} , 
    && \mathbb{M}^{\rho^\spb} = \begin{pmatrix}
        0 & -\mathcal{D}^\spb q^2 & 0 & i v_0^\spb (\hat{\bm u}_0 \cdot \bm q) & 0 & i v_0^\spb (\hat{\bm u}_{\perp,0} \cdot \bm q)
    \end{pmatrix} , 
\end{align*}
\begin{align*}
%%%%%%%%%%%%%%%%%%%%%%%%%%%%%%
    & \ \mathbb{M}^{s^\sa} = \begin{pmatrix}
        \frac{v_0^\sa}{2} i q_x + s_0^\sa f_{\sa,\rho^\sa}^{\sa} 
        +\varepsilon s_0^\spb f_{\spb,\rho^\sa}^{\sa} \\
        s_0^\sa f_{\sa,\rho^\spb}^{\sa}
        + \varepsilon s_0^\spb f_{\spb,\rho^\spb}^{\sa} \\
        i (\bm K_{\sa,0}^\sa \cdot \bm q) + f_{\sa,0}^{\sa} + s^\sa_0 f_{\sa,s^\sa}^{\sa}
        + \varepsilon s^\spb_0 f_{\spb,s^\sa}^{\sa}
        - q^2 D^\sa_\sa \\
        i (\bm K_{\spb,0}^\sa \cdot \bm q) 
        + \varepsilon f_{\spb,0}^{\sa} + s^\sa_0 f_{\sa,s^\spb}^{\sa}
        + \varepsilon s^\spb_0 f_{\spb,s^\spb}^{\sa}
        - \varepsilon q^2 D^\sa_\spb \\
        i (\bm L_{\sa,0}^\sa \cdot \bm q) 
        + f_{\sa,\theta^\sa}^{\sa}
        + \varepsilon\frac{s_0^{\spb}}{s_0^\sa} f_{\spb,\theta^\sa}^{\sa} \\
        i (\bm L_{\spb,0}^\sa \cdot \bm q) 
        + \frac{s_0^{\sa}}{s_0^\spb} f_{\sa,\theta^\spb}^{\sa}
        +\varepsilon f_{\spb,\theta^\spb}^{\sa}
    \end{pmatrix}^T , &&
    &  \mathbb{M}^{\theta^\sa} = \begin{pmatrix}
        \frac{v_0^\sa}{2} i q_y  \\
      0\\
        i (-\bm L_{\sa,0}^\sa \cdot \bm q)  \\
        i (-\bm L_{\spb,0}^\sa \cdot \bm q) \\
        i (\bm K_{\sa,0}^\sa \cdot \bm q) 
        - \varepsilon \frac{s_0^{\spb}}{s_0^{\sa}} f_{\spb,0}^{\sa} 
        - q^2 D^\sa_\sa \\
        i (\bm K_{\spb,0}^\sa \cdot \bm q) 
        + \varepsilon f_{\spb,0}^{\sa} 
        - q^2 \varepsilon D^\sa_\spb 
    \end{pmatrix}^T ,
\end{align*}
\begin{align*} 
%%%%%%%%%%%%%%%%%%%%%%%%%%%%%%
 & \mathbb{M}^{s^\spb} = \begin{pmatrix}
           \varepsilon s_0^\sa f_{\sa,\rho^\sa}^{\spb}
        + s_0^\spb f_{\spb,\rho^\sa}^{\spb} \\ \frac{v_0^\spb}{2} i (\hat{\bm u}_0 \cdot \bm q) + 
       \varepsilon s_0^\sa f_{\sa,\rho^\spb}^{\spb}
        + s_0^\spb f_{\spb,\rho^\spb}^{\spb} \\
        i (\bm K_{\sa,0}^\spb \cdot \bm q) 
        + \varepsilon f_{\sa,0}^{\spb} 
        + \varepsilon s^\sa_0 f_{\sa,s^\sa}^{\spb}
        + s^\spb_0 f_{\spb,s^\sa}^{\spb} 
        - \varepsilon q^2 D^\spb_\sa \\
        i (\bm K_{\spb,0}^\spb \cdot \bm q) 
        + f_{\spb,0}^{\spb} 
        + \varepsilon s^\sa_0 f_{\sa,s^\spb}^{\spb}
        + s^\spb_0 f_{\spb,s^\spb}^{\spb}
        - q^2 D^\spb_\spb \\
        i (\bm L_{\sa,0}^\spb \cdot \bm q) 
        + \varepsilon f_{\sa,\theta^\sa}^{\spb}
        + \frac{s^\spb_0}{s_0^\sa}f_{\spb,\theta^\sa}^{\spb} \\
        i (\bm L_{\spb,0}^\spb \cdot \bm q) 
        + \varepsilon\frac{s_0^{\sa}}{s_0^\spb} f_{\sa,\theta^\spb}^{\spb}
        + f_{\spb,\theta^\spb}^{\spb}
    \end{pmatrix}^T , 
 &  
 %%%%%%%%%%%%%%%%%%%%%%%%%%%%%%
    &&\mathbb{M}^{\theta^\spb} = \begin{pmatrix}
        0 \\
        \frac{v_0^\spb}{2} i (\hat{\bm u}_{\perp,0} \cdot \bm q) \\
        i (-\bm L_{\sa,0}^\spb \cdot \bm q) \\
        i (-\bm L_{\spb,0}^\spb \cdot \bm q) \\
        i (\bm K_{\sa,0}^\spb \cdot \bm q) 
        + \varepsilon f_{\sa,0}^{\spb} - q^2 \varepsilon D^\spb_\sa \\
        i (\bm K_{\spb,0}^\spb \cdot \bm q) 
        - \varepsilon\frac{s_0^{\sa}}{s_0^{\spb}} f_{\sa,0}^{\spb} 
        - q^2 D^\spb_\spb 
    \end{pmatrix}^T.
\end{align*}
Here, the advection terms are given by
\begin{align*}
    (\bm K_{\sa,0}^\sa \cdot \bm q)  & = \left[ s^\sa_0\psi^\sa_{\sa\sa} 
    + s^\spb_0\psi^\sa_{\sa\spb} \varepsilon
    \right] q_x, &&  (\bm L_{\sa,0}^\sa \cdot \bm q) =  \left[ s^\sa_0\lambda^\sa_{\sa\sa} 
    + s^\spb_0 \varepsilon \lambda^\sa_{\spb\sa} \right] q_y  \\
    (\bm K_{\spb,0}^\sa \cdot \bm q) & = 
    \left[ s^\sa_0 \varepsilon \psi^\sa_{\spb\sa}
    + s^\spb_0 \psi^\sa_{\spb\spb}\right] q_x, && (\bm L_{\spb,0}^\sa \cdot \bm q)  = 
    \left[ s^\sa_0 \varepsilon \lambda^\sa_{\sa\spb} 
    + s^\spb_0 \lambda^\sa_{\spb\spb}\right] q_y ,  \\
    (\bm K_{\sa,0}^\spb \cdot \bm q) & = 
    \left[ s^\sa_0 \varepsilon \psi^\spb_{\sa\sa}
    + s^\spb_0  \psi^\spb_{\sa\spb}\right]q_x && (\bm L_{\sa,0}^\spb \cdot \bm q) 
    =  \left[ s^\sa_0 \varepsilon \lambda^\spb_{\sa\sa} 
    + s^\spb_0 \lambda^\spb_{\spb\sa}
    \right] q_y, \\
    (\bm K_{\spb,0}^\spb \cdot \bm q) & = \left[ s^\sa_0\psi^\spb_{\spb\sa} 
    + s^\spb_0 \varepsilon \psi^\spb_{\spb\spb} \right] q_x, &&  (\bm L_{\spb,0}^\spb \cdot \bm q) = \left[ 
    s^\sa_0 \lambda^\spb_{\sa\spb}
    + s^\spb_0 \varepsilon  \lambda^\spb_{\spb\spb}
    \right] q_y ,\\
\end{align*}
where $\varepsilon = \cos\Delta\theta_{\sa\spb} = + 1$ for the homogeneous flocking solution and $-1$ for the homogeneous anti-flocking solution.

\subsection{Robustness to population heterogeneity}

\begin{figure}[h!]
    \centering
    \includegraphics[width=0.7\linewidth]{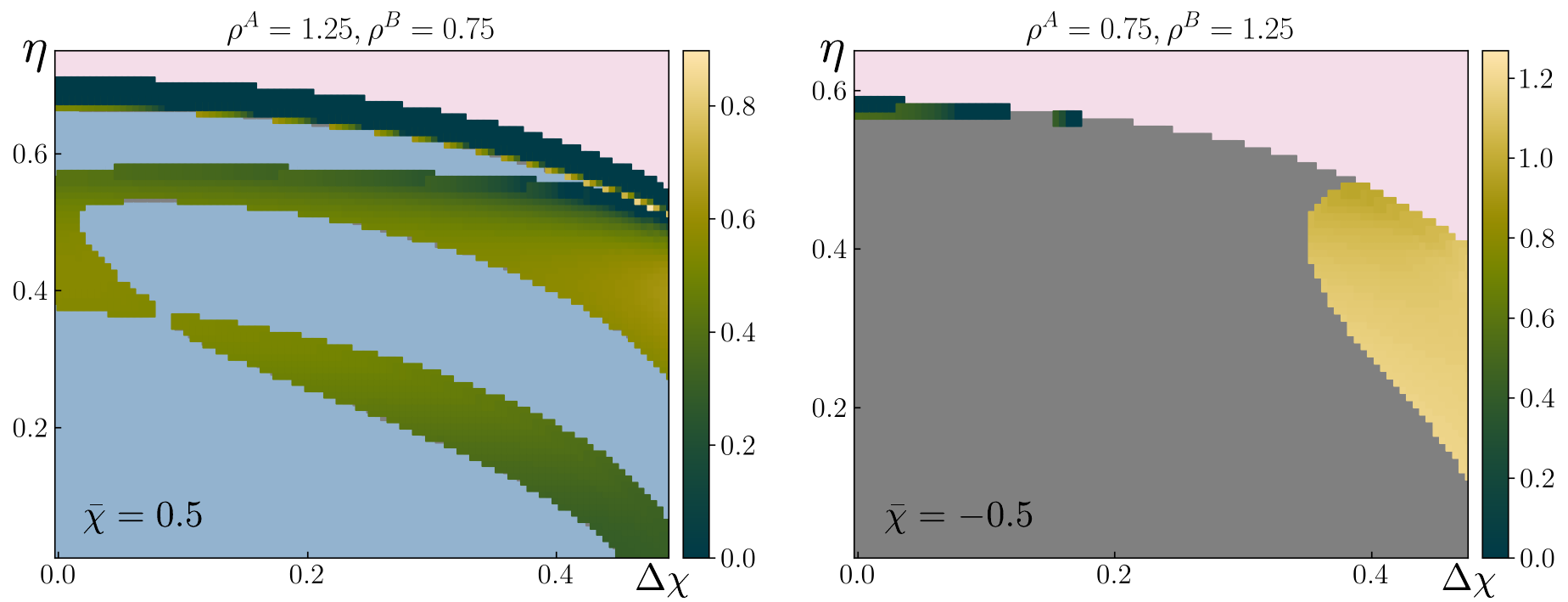}
    \caption{Phase diagrams in the $(\Delta\chi, \eta)$ plane obtained from the linear stability of the stationary homogeneous solutions of Eqs. (2) in the flocking (a) and anti-flocking (b) regimes for unequal species densities. Stable disordered, flocking and anti-flocking regions appear in pink, blue and gray, respectively. The other regions indicate an instability, and are colored according to the orientation of the most unstable wave vector with respect to the direction of order.}
    \label{fig:figS6}
\end{figure}

In the main text, we focus on the sole induce of nonreciprocity.
Therefore, we considered two species with the same intrinsic parameters ($\eta = \eta^{\sa,\spb}, v_0 = v_0^{\sa,\spb}$ and $\rho_0 = \rho^{\sa,\spb}$). Nevertheless, our analytical results suggest that nonreciprocal interactions generically destabilise nonreciprocal flocks, and thus that the resulting instability is robust to population heterogeneity. To substantiate this claim, we show in Fig.~\ref{fig:figS6} phase diagrams analogous to that of Figure~3 in the main text, but corresponding to unequal mean densities $\rho^\sa \neq \rho^\spb$ of the two species. 
As in the equal density case, both phase diagrams exhibit an instability within the homogeneous ordered phase associated with a growth rate $\Lambda \sim |\bm q|$ (not shown) in the limit of small wavenumbers. 
Hence, this data confirms that demixing of flocks induced by weak non-reciprocity should remain robust to the presence of population heterogeneity.

\subsection{Enslaving of the non-hydrodynamic modes}

 Here, we detail the enslaving procedure leading to Eq.~(4) of the main text. We first outline the general procedure to identify the non-hydrodynamic modes and to enslave them to the hydrodynamic modes. We then apply this formalism to our specific system. \\
 
 We consider a $m$-dimensional linear system of the form $\partial_t \bm X = \mathbb{M}(\bm q) \cdot \bm X$, where $\mathbb{M}(\bm q)$ is a wave-vector dependent $m\times m$ matrix. 
As we are interested in the $q = |\bm q| \to 0$ limit, we expand $\mathbb{M}$ 
as 
\begin{equation} \label{sm_Mexp}
    \mathbb{M}(\bm q) = \mathbb{M}_0 + i q \mathbb{M}_1 - q^2 \mathbb{M}_2 + \mathcal{O}(q^3),
\end{equation}
where the $\mathbb{M}_i$ matrices are independent of $q$ but may depend on the direction of $\bm q$.

We denote $\{\bm Y_{j}\}$ as the left eigenbasis of $\mathbb{M}_0$, such that for all $j = 1, \ldots, m$, $\bm Y_j^T \cdot \mathbb{M}_0 = \Lambda_j \bm Y_j^T$,
where $\{\Lambda_j\}$ are the corresponding eigenvalues.
Assuming that the initial linear system describes fluctuations around a solution which is stable under $q = 0$ perturbations, all $\Lambda_j \le 0$.
We now suppose that the problem has $n_{\rm h}$ hydrodynamic modes, and we reorder the eigenvalues such that
$\Lambda_1, \ldots, \Lambda_{n_{\rm h}} = 0$ and $\Lambda_{n_{\rm h}+1}, \ldots, \Lambda_{m} < 0$.
Defining 
\begin{equation} \label{sm_tildeX}
    \Tilde{\bm X} = \begin{pmatrix} \bm Y_1 & \cdots & \bm Y_m\end{pmatrix}^T \cdot \bm X \equiv \mathbb{P} \cdot \bm X,
\end{equation}
the first $n_{\rm h}$ elements of $\Tilde{\bm X}$ are the hydrodynamic modes, while the other $m -n_{\rm h}$ are fast.
This can be straightforwardly shown from
\begin{align}
    \partial_t \Tilde{\bm X} & \underset {q = 0}{=} \mathbb{P} \cdot \partial_t \bm X = \mathbb{P} \cdot \mathbb{M}_0 \cdot \bm X = {\rm diag}(\lambda_1 \dots \lambda_m)\cdot \mathbb{P} \cdot \bm X = {\rm diag}(\lambda_1 \dots \lambda_m) \cdot \Tilde{\bm X},
\end{align}
where we have used that $\mathbb{P} \cdot \mathbb{M}_0 \cdot \mathbb{P}^{-1} = {\rm diag}(\lambda_1 \dots \lambda_m)$.
While the slow modes are uniquely defined, the fast modes may be expressed differently so long as they are associated with a nonzero damping rate. The formulation~\eqref{sm_tildeX} nevertheless provides a compact change of basis formula between the slow and fast modes and the original variables.
Going back to $q > 0$, the dynamics of $\Tilde{\bm X}$ then follows
$\partial_t\Tilde{\bm X} = \Tilde{\mathbb{M}}(\bm q)\cdot \Tilde{\bm X}$, where $\Tilde{\mathbb{M}} \equiv \mathbb{P}\cdot \mathbb{M} \cdot \mathbb{P}^{-1}$. 
In the regime of interest, $\mathbb{P}$ is always invertible. 
For convenience, we write separately the dynamics of the hydrodynamic ($\bm X_{\rm h}$) and fast ($\bm X_{\rm f}$) modes (dropping the tildes to lighten notations):
\begin{align}
        \partial_t\bm X_{\rm h} - \mathbb{M}^{\rm h}_{\rm h} \cdot \bm X_{\rm h} 
        & = \mathbb{M}^{\rm h}_{\rm f} \cdot \bm X_{\rm f}  ,\label{evol_slow_modes}\\
        \partial_t\bm X_{\rm f}  - \mathbb{M}^{\rm f} _{\rm f}  \cdot \bm X_{\rm f} 
        & = \mathbb{M}^{\rm f}_{\rm h} \cdot \bm X_{\rm h},\label{evol_fast_modes}
\end{align}
in terms of the block matrices $\mathbb{M}^x_y$.
By construction, all elements of $\mathbb{M}^{\rm h}_{\rm h}$ and $\mathbb{M}^{\rm h}_{\rm f}$ are zero for $q = 0$, while $\mathbb{M}^{\rm f}_{\rm f}(q=0)$ is diagonal with negative entries representing the finite damping rates of the fast modes. The time evolution of the fast modes is obtained by solving the PDE~\eqref{evol_fast_modes}, 
\begin{align*}
    \bm X_{\rm f} &= \int_{0}^{t} e^{\mathbb{M}^{\rm f}_{\rm f}(t-t')}\cdot \mathbb{M}^{\rm f}_{\rm h} \cdot \bm X_{\rm h}(\bm q,t') \, d t' + e^{\mathbb{M}_{\rm f}^{\rm f}(t-t_0)}\cdot\bm X_{\rm f}(0).
\end{align*} 
In the long-time limit $t \to \infty$, the second term can be neglected, as $t$ is much larger than any characteristic damping timescale associated with $\mathbb{M}^{\rm f}_{\rm f}$. Performing the change of variables $\tau = t - t'$ then yields
\begin{align*}
    \bm X_{\rm f} &= \int_{0}^{\infty} e^{\mathbb{M}^{\rm f}_{\rm f}\, \tau} \cdot \mathbb{M}^{\rm f}_{\rm h} \cdot \bm X_{\rm h}(\bm q,t - \tau) \, {\rm d} \tau \\
    &= \int_{0}^{\infty} e^{\mathbb{M}^{\rm f}_{\rm f} \, \tau} \cdot \mathbb{M}^{\rm f}_{\rm h} \cdot (1 - \tau \partial_t + \frac{\tau^2}{2}\partial^2_{tt} + \dots) \bm X_{\rm h}(\bm q,t) {\rm d}\tau \\
    & = - (\mathbb{M}^{\rm f}_{\rm f})^{-1} \cdot \mathbb{M}^{\rm f}_{\rm h} \cdot  \bm X_{\rm h}(\bm q,t) + ((\mathbb{M}^{\rm f}
    _{\rm f})^{-1})^2 \cdot \mathbb{M}^{\rm f}_{\rm h} \cdot \partial_t \bm X_{\rm h}(\bm q,t) + \mathcal{O}(\tau^2),
\end{align*}
where, in the last step, we have used integration by parts. Expanding all block matrices similarly to~\eqref{sm_Mexp}, and using that $(\mathbb{M}^{\rm x}_{\rm x})^{-1} = (\mathbb{M}^{\rm x}_{\rm x,0} + iq \mathbb{M}^{\rm x}_{\rm x,1} + \mathcal{O}( q^2))^{-1} = (\mathbb{M}^{\rm x}_{\rm x,0})^{-1} - i q(\mathbb{M}^{\rm x}_{\rm x,0})^{-1}\cdot \mathbb{M}^{\rm x}_{\rm x,1}\cdot (\mathbb{M}^{\rm x}_{\rm x,0})^{-1} + \mathcal{O}(q^2)$, we find that in the limit $q \to 0$, the second term on the r.h.s.\ of the above equality vanishes, while the first term gives 
\begin{align*}
    \bm X_{\rm f} = -\left({\mathbb{M}}^{\rm f}_{{\rm f},0}\right)^{-1}\cdot {\mathbb{M}}^{\rm f}_{{\rm h},0} \cdot \bm X_{\rm h} + \mathcal{O}(q).
\end{align*}
We then enslave the fast modes by substituting this expression in Eq.~\eqref{evol_slow_modes},
and finally find that the hydrodynamic modes obey $\partial_t\bm X_{\rm h} = \mathbb{M}_{\rm h}(\bm q) \cdot \bm X_{\rm h}$ with
\begin{equation}
    \mathbb{M}_{\rm h}(\bm q) = iq\left( \mathbb{M}^{\rm h}_{{\rm h},1} - \mathbb{M}^{\rm h}_{{\rm f},1}\cdot \left({\mathbb{M}}^{\rm f}_{{\rm f},0}\right)^{-1}\cdot {\mathbb{M}}^{\rm f}_{{\rm h},0} \right) + {\cal O}(q^2).
\end{equation}

Performing this analysis for the linear stability described in the main text, we find three vanishing left eigenvalues of $\mathbb{M}_0$ corresponding to the slow modes: $\delta\hat{\rho}^{\sa,\spb}$ and $\bar{\theta} = s^\sa_0 F^{\spb}_{\sa}\delta \hat{\theta}^\sa + {s^\spb_0}{F^{\sa}_{\spb}}\delta\hat{\theta}^\spb$.
The expressions of the various matrices involved in the derivation are quite lengthy and not very informative.
We can nevertheless express the coefficients that couple the density fluctuations in Eq.~(4) of the main text as
\begin{align*}
    V_{\sa\sa} & = v_0^{\sa} \left[ C_{\sa\sa} E^{\sa}_{\sa} + C_{\sa\spb} E^{\spb}_{\sa} \right], &
    V_{\sa\spb} & = v_0^{\sa} \left[ C_{\sa\sa} E^{\sa}_{\spb} + C_{\sa\spb} E^{\spb}_{\spb} \right],\\
    V_{\spb\sa} & = v_0^{\spb} \left[ C_{\spb\sa} E^{\sa}_{\sa} + C_{\spb\spb} E^{\spb}_{\sa} \right], &
    V_{\spb\spb} & = v_0^{\spb} \left[ C_{\spb\sa} E^{\sa}_{\spb} + C_{\spb\spb} E^{\spb}_{\spb} \right],
\end{align*}
where the $C_{\sps\su}$ coefficients are constant and depend on the model parameters, while $E^{\sps}_{\su} = \left(s_0^{\sa}\partial_{\rho^{\su}}\alpha^{\sps}_{\sa} + \epsilon s_0^{\spb}\partial_{\rho^{\su}}\alpha^{\sps}_{\spb}\right)_{\rho^{\sps} = \rho_0}$ with $\epsilon = {\rm sign}(\bar{\chi})$.

\section{Description of the supplemental movies}

All movies are composed as follows: the top row shows each particle coloured according to their species (left) and orientation (right). The caption is the same as that given in Fig.~1 of the main text. The bottom row shows time series of both species order parameter norms (left) and orientation (right, orientations are normalized by $\pi$). The color code follows that of the species, red for $\sa$ and blue for $\spb$.
In all movies we use: $\rho_0 = \frac{1}{2}$, $v_0 = 1$ and $\bar\chi = -\frac{1}{2}$.
\begin{itemize}
    \item \textbf{SMov~1:} Dyanmics of the laning pattern over short timescales ($10^4$ timesteps). The lanes form perpendicular to the direction of propagation. Simulation parameters are $\Delta\chi = 0.35, \eta = 0.3$ and $L= 512$.  
    \item \textbf{SMov~2:} Dynamics of the laning pattern over longer timescales ($10^5$ timesteps). The movie illustrates that this pattern is highly dynamic, with lanes that persistently rotate, as reflected by the global polarization angle $\Phi^\sps(t)$, and drift. Simulation parameters are the same as in SMov~1.   
    \item \textbf{SMov~3:} Dynamics in the chaotic cluster phase, where the two species form polar clusters predominantly composed of one species. These clusters move in a chaotic manner and lack macroscopic spatial order. The timescale ($10^4$ timesteps) is identical to the one used in SMov~1, allowing direct comparison. Simulation parameters are $\Delta\chi = 0.45, \eta = 0.3$ and $L= 512$.  
    \item \textbf{SMov~4:} Movie illustrating lane-breaking events, which are associated with a sharp decay of polar order. Simulation parameters are the same as in SMov~2, except for the system size which is here $L = 2048$, since lane-breaking events occur more frequently in larger systems. 
\end{itemize}

Supplemental movies and a Mathematica notebook allowing to compute the coefficients of the continuum model are available at: \url{https://owncloud.gwdg.de/index.php/s/JSCGDzh2vRj3ETs}.